\documentclass[aps,twocolumn,prb,superscriptaddress]{revtex4-2}

\usepackage[pdftex]{graphicx,hyperref}
\usepackage[utf8]{inputenc}
\usepackage{amsmath,bm}
\usepackage{color,soul}
\usepackage{xstring}
\usepackage{wasysym}
\usepackage{xspace}
\usepackage{amssymb}
\usepackage{time}
\usepackage{subfigure}
\usepackage{multirow}
\usepackage{xspace}
\usepackage{url}
\usepackage[normalem]{ulem}

\def\<{\langle}
\def\>{\rangle}
\DeclareMathOperator{\Tr}{Tr}

\newcommand{\ve}[1]{\boldsymbol{#1}}

\makeatletter
\def\maketitle{
\@author@finish
\title@column\titleblock@produce
\suppressfloats[t]}

\begin{document}
\title{Fermion disorder operator at Gross-Neveu and deconfined quantum criticalities}
\author{Zi Hong Liu}
\affiliation{Institut für Theoretische Physik und Astrophysik and Würzburg-Dresden Cluster of Excellence ct.qmat,
	Universität Würzburg, 97074 Würzburg, Germany}

\author{Weilun Jiang}
\affiliation{Beijing National Laboratory for Condensed Matter Physics and Institute of Physics, Chinese Academy of Sciences, Beijing 100190, China}
\affiliation{School of Physical Sciences, University of Chinese Academy of Sciences, Beijing 100190, China}

\author{Bin-Bin Chen}
\affiliation{Department of Physics and HKU-UCAS Joint Institute
	of Theoretical and Computational Physics, The University of Hong Kong,
	Pokfulam Road, Hong Kong SAR, China}

\author{Junchen Rong}
\affiliation{Institut des Hautes Études Scientifiques, 91440 Bures-sur-Yvette, France}

\author{Meng Cheng}
\email{m.cheng@yale.edu}
\affiliation{Department of Physics, Yale University, New Haven, CT 06520-8120, USA}

\author{Kai Sun}
\email{sunkai@umich.edu}
\affiliation{Department of Physics, University of Michigan, Ann Arbor, Michigan 48109, USA}

\author{Zi Yang Meng}
\email{zymeng@hku.hk}
\affiliation{Department of Physics and HKU-UCAS Joint Institute
	of Theoretical and Computational Physics, The University of Hong Kong,
	Pokfulam Road, Hong Kong SAR, China}

\author{Fakher F. Assaad}
\email{fakher.assaad@physik.uni-wuerzburg.de}
\affiliation{Institut für Theoretische Physik und Astrophysik and Würzburg-Dresden Cluster of Excellence ct.qmat,
	Universität Würzburg, 97074 Würzburg, Germany}
	
\begin{abstract}
The fermion disorder operator has been shown to reveal the entanglement information in 1D Luttinger liquids and 2D free and interacting Fermi and non-Fermi liquids emerging  at quantum critical points (QCP)~\cite{jiangFermion2022}.  Here we study, by means of large-scale quantum Monte Carlo simulation, the scaling behavior of disorder operator in correlated Dirac systems. We first demonstrate the logarithmic scaling behavior of the  disorder operator at the Gross-Neveu (GN) chiral Ising and Heisenberg QCPs,  where consistent conformal field theory (CFT) content of the GN-QCP in its coefficient is found. Then we study a 2D  monopole free  deconfined quantum critical point (DQCP) realized between a quantum-spin Hall insulator and a superconductor.  Our  data  point  to  negative  values of the logarithmic coefficients  such  that the DQCP does not  correspond  to a unitary CFT.
Density matrix renormalization group calculations of the  disorder operator on a 1D DQCP model also detect emergent continuous symmetries. 
\end{abstract}
\date{\today}
\maketitle

\noindent{\textcolor{blue}{\it Introduction and Motivation.}---} 
Entanglement  witnesses  can reveal the fundamental organizing principle of many-body systems. { One of such witness is the disorder  operator  \cite{Kadanoff1971,Fradkin2017,Nussinov2006,Nussinov2009}, which hinges   on    symmetry  
properties of   a  system,  such as  spin rotational    invariance  or  charge  conservation. It is an equal-time observable with no need for a replica manifold  and  is  easily  accessible  to  auxiliary field 
determinantal  quantum Monte  Carlo  (DQMC) simulations~\cite{Blankenbecler81,White89,Assaad08_rev}. 
For a given system with a U(1) global symmetry, with generator $\hat{Q}=\sum_{i\in V}\hat{n}_{i}$, the disorder operator carries  out a  rotation with angle $\theta$ in the entanglement region $M \subset  V $, i.e., $X\left(\theta\right)=\langle  \prod_{i\in M}\exp\left(i\theta\hat{n}_{i}\right) \rangle $, as shown schematically in Fig.~\ref{fig:qsh-phase-diagram} (d) and (e).}  For states of matter   characterized  by  a  finite length scale,    such as  a band insulator,  this  rotation  only  effects  the  boundary,      and   generically   
an area  law is   expected.   For  scale  invariant  systems  logarithmic corrections to  the  area  law reveal  critical  behavior \cite{jiangFermion2022,zhaoHigher2021,wangScaling2021,wang2021ScalingDQCP} .  Furthermore,  sub-leading   corrections  reflect  topological order  \cite{chen2022topological}. 
 { In particular at a (conformal) QCP, when the boundary of region $M$ is not smooth, the scaling behavior
of the disorder operator -- {similar with that of the EE} -- acquires logarithmic corner correction term~\cite{Fradkin2006} 
\begin{equation}
\ln \left|X(\theta)\right| 
\sim -al+s(\theta)\ln l+c.  
	\label{Xscaling}
\end{equation}
While  the   area  law  coefficient $a$ is sensitive to the UV  physics,   the  log coefficient $s(\theta)$  is universal and reflects  the IR physics. }

The  above   has  an  obvious  overlap  with  the entanglement entropy (EE)  and  related  entanglement spectrum (ES)~\cite{CARDY1988,Srednicki1993,Christoph1994,Calabrese_2004,Fradkin2006,Casini2006,Kitaev2006top,Levin2006,liEntanglement2008,poilblanc2010entanglement,FSong2012,grover2013entanglement,Assaad2014,Assaad2015,laflorencieQuantum2016,Parisen2018,dEmidio2020Entanglement,JRZhao2022,yanRelating2021,JRZhao2022QiuKu,demidioUniversal2022,songReversing2022,liaoTeaching2023,panComputing2023}. Although in  some   special  cases  both the {$n$-th order R\'enyi EE, i.e., $-\frac{1}{1-n}\ln \Tr (\rho_M^n)$ where $\rho_M=\mathrm{Tr}_{\overline{M}}\rho$ is the reduced density matrix} and  the  disorder  operator   produce  identical  results,   both  quantities  differ.    Being  symmetry  based  the  disorder  operator   offers  more  possibilities  such   as  
the  detection of  emergent  symmetries.  The  aim  of  this letter  is  to  investigate  these  possibilities     for Dirac  systems.

As  mentioned above,  the  disorder operator and EEs are different quantities, with the former formulated  in terms of a  global  symmetry of   the model  system and the latter defined  without any  symmetry considerations. However, their connections from entanglement perspective can be established from two important limits of the disorder operator. The first one is in the
small angle limit $\theta\rightarrow0$, where the disorder operator can
be mapped to bipartite fluctuations $- \ln \left|X(\theta)\right|\propto  \theta^{2}   \langle \left( \sum_{i \in M } (\hat{n}_{i} - \langle \hat n_i \rangle) \right)^2 \rangle $~\cite{song2012bipartite}.
  In the small $\theta$ limit, $s(\theta)$ is proportional to the central charge in the conformal field theory (CFT) critical point~\cite{zhaoHigher2021,wangScaling2021,wang2021ScalingDQCP,jiangFermion2022}. The second limit is for non-interacting systems~\cite{zhaoHigher2021, jiangFermion2022}.  For fermions the 
disorder operator at special angles maps onto the R\'enyi EEs, for example $S_2=-2 \ln|X(\pi/2)|$ and $S_3 =-\ln \left|X(2\pi/3)\right|$. The general case can be found in \cite{jiangFermion2022} and a similar relation holds for the  free scalar
theory~\cite{zhaoHigher2021}.  

Beyond the Gaussian limit,  one needs to study the disorder operator   with  numerical simulations. For  boson/spin and topologically ordered systems, it was shown that 
 the log-coefficient  reflects the CFT content of the QCP~\cite{wangScaling2021,zhaoHigher2021} and the remaining constant reflects the topological degeneracy~\cite{chen2022topological}. Of special importance to  this paper,  is the disorder operator of the 2D JQ spin model~\cite{sandvikEvidence2007} for the  deconfined  QCP (DQCP)~\cite{senthilQuantum2004}  where   the log-coefficient at $\theta=\pi$ is  negative~\cite{wang2021ScalingDQCP}. This  finding, together with the similar negative log-coefficient in  $S_2$~\cite{JRZhao2022}, suggests that the JQ model realization of the DQCP may not be an unitary CFT (or not a CFT at the first place)~\cite{Casini2012,liaoTeaching2023}, and calls  to reconsider  DQCP related phenomena such as the emergent  symmetries~\cite{sandvikEvidence2007,maRole2019},  dangerously irrelevant operators~\cite{Shao2016,Nahum2015}, weakly first order transition~\cite{kuklovDeconfined2008,chenDeconfined2013,emidioDiagnosing2021}, complex fixed points and walking of scaling dimensions~\cite{nahumNote2020,maTheory2020,gorbenkoWalking2018},  multicriticalities~\cite{zhaoMulticritical2020}, etc, and the physical mechanism behind them.

Hence, the question arises if  the negative log-coefficient observed at the JQ model realization of DQCP is a generic feature or an artifact of the model. Therefore, it is of importance to measure the disorder operator at other 2D DQCP lattice models and  also in 1D systems~\cite{huangEmergent2019,robertsDeconfined2019}. The cleanest one present, without any difficulties/ambiguities of the two length scale~\cite{Shao2016,Nahum2015}, is the model based on the interacting Dirac fermions~\cite{liu2019superconductivity,liuGross2021} with emergent SO(5) symmetry from order parameter measurements, as shown in Fig.~\ref{fig:qsh-phase-diagram} (a) and Eq.~\eqref{eq:model}.  It is the focus of this paper to compute the disorder operator at this DQCP in large scale DQMC simulations. 

\begin{figure}[htp!]
	\includegraphics[width=\columnwidth]{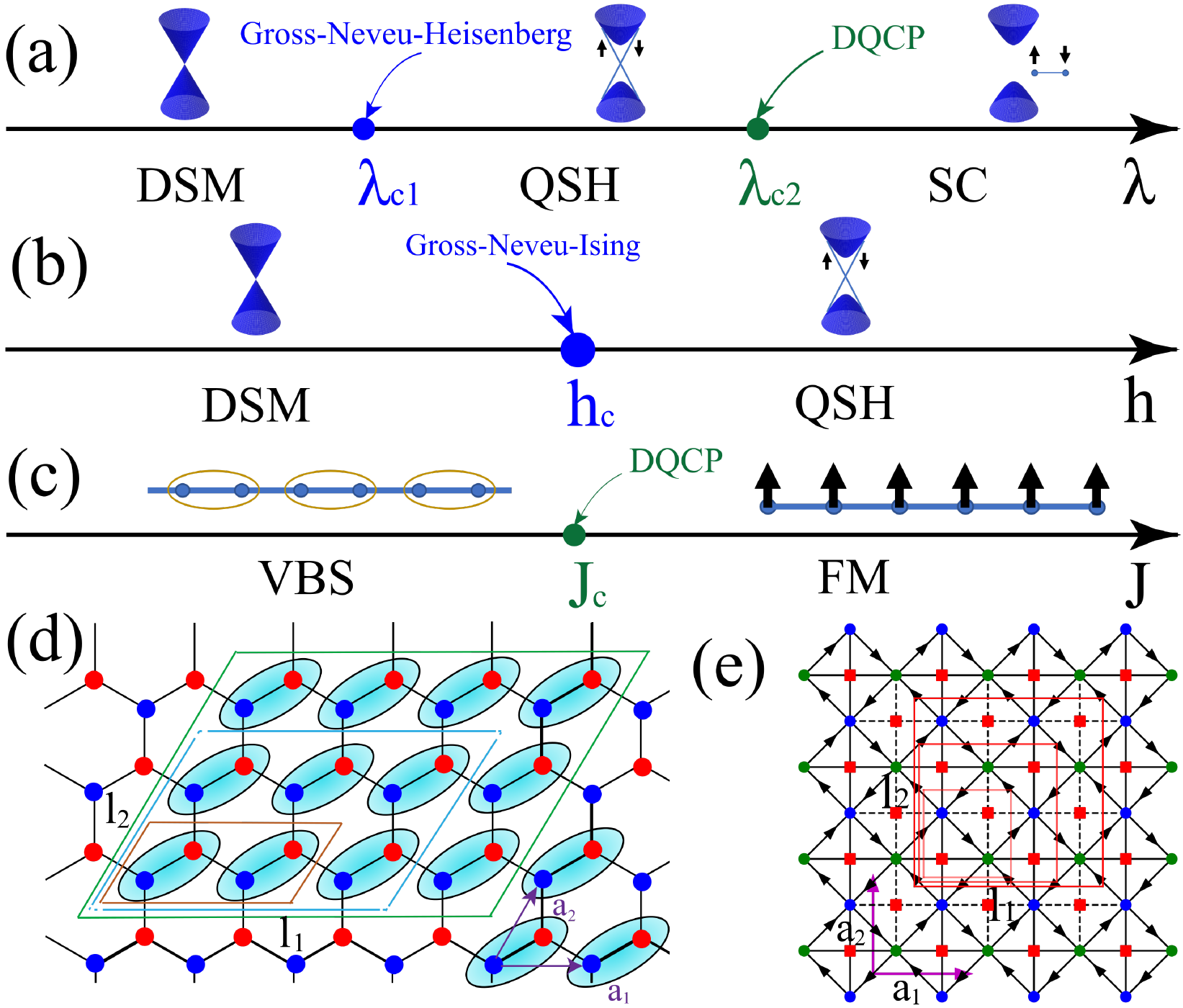}
	\caption{(a,b,c) Schematic phase diagrams of the 2D DQCP model in Eq.~\eqref{eq:model}, the $\pi$-flux model and the 1D DQCP model. The 2D DQCP model has three phases: DSM, QSH, and SC, separated by two QCPs: GN-Heisenberg QCP at $\lambda_{c1}$ and DQCP at $\lambda_{c2}$~\cite{liu2019superconductivity}. For comparison, the $\pi$-flux model also possesses DSM and QSH phases, separated by the GN-Ising QCP at $h_c$~\cite{he2018Dynamical,liuDesigner2020,wangEmus2023}. The last one is a 1D model containing VBS and FM phases separated by a DQCP, with  $[\mathrm{U}(1)\times\mathrm{U}(1)]\rtimes\mathbb{Z}_2$ emergent  symmetry~\cite{huangEmergent2019,robertsDeconfined2019}. (d,e) Sketch of parallelogram entanglement region $M$ in the 2D DQCP model and square entanglement region $M$ in the $\pi$-flux model. $\mathbf{a}_1, \mathbf{a}_2$ are unit vectors, and $l_1$ and $l_2$ are the linear lengths of $M$. We define the perimeter $l = 2(l_1 + l_2) $ for both models,  which is used to extract the scaling behavior of the disorder operators.}
	\label{fig:qsh-phase-diagram}
\end{figure}


\noindent{\textcolor{blue}{\it Models and numerical settings.}---}
We study the model introduced in Ref.~\cite{liu2019superconductivity} with Hamiltonian
\begin{equation}
\hat H = -t\sum_{\left\langle ij\right\rangle }(\hat{\ve{c}}_{i}^{\dagger}\hat{\ve{c}}_{j}+\mathrm{h.c.})-\lambda\sum_{\hexagon} \Bigg( \sum_{ \langle \langle \ve{i},\ve{j} \rangle \rangle  \in \hexagon } \mathrm{i}  \nu_{\boldsymbol{ij}}
 \hat{\boldsymbol{c}}^{\dagger}_{\ve{i}} \boldsymbol{\sigma} \hat{\boldsymbol{c}}^{}_{\ve{j}}+\text{H.c.} \Bigg)^2\label{eq:model}
\end{equation}
where $\left\langle \cdots\right\rangle $ and $\left\langle \left\langle \cdots\right\rangle \right\rangle $
refer to  nearest and next nearest  neighbors on  the honeycomb lattice (as shown in Fig.~\ref{fig:qsh-phase-diagram} (d)).
The phase diagram has been mapped out with DQMC simulations~\cite{liu2019superconductivity,liuGross2021,ALFSciPost}, see Fig.~\ref{fig:qsh-phase-diagram} (a). As a
function of a single parameter $\lambda$ (with $t=1$ as the energy unit), the phase diagram shows
a GN-Heisenberg QCP at $\lambda_{c1}=0.0187(2)$ separating a Dirac semi-metall (DSM)  and a  quantum spin Hall (QSH) insulator,  
and a  DQCP at $\lambda_{c2}=0.0332(2)$ separating the QSH and s-wave superconductor (SC). 
%
%
The key difference  between Eq.~\eqref{eq:model} and the JQ model  is  the  absence of  monopoles.  In the  JQ  model,  the  U(1)  symmetry  
is  emergent  since the lattice breaks  it  down  to  $\mathbb{Z}_4$.  As  a  consequence,  quadruple  monopoles  are  symmetry  allowed,  and  a  second length scale   at  which  the  $\mathbb{Z}_4$  symmetry  is enhanced to U(1)   rotational  symmetry  
obscures the numerical analysis~\cite{Shao2016}.  In the model Eq.~\eqref{eq:model},  the U(1) symmetry   corresponds  to charge conservation present in the microscopic model {(we note the same microscopic U(1) symmetry also occurs in a closely related cubic dimer model~\cite{sreejithEmergent2019}).} Although the computational complexity of DQMC for Dirac fermion is much higher than that of the stochastic series expansion QMC~\cite{sandvikEvidence2007} for the JQ model,    order  parameter  calculations   support   that   SU(2) spin  and  U(1)  charge  symmetries are enhanced to an  emergent  SO(5) symmetry at this DQCP.  As we shall see below, we still find the non-unitary signature of the DQCP in this model, similar to that of the JQ model~\cite{wang2021ScalingDQCP,JRZhao2022}. {We also find that the logarithmic corrections of disorder operators appear to violate the emergent SO(5) symmetry.}

To probe the  U(1) charge  and the SU(2)  spin  symmetries,  we consider the disorder operators 
\begin{equation}
X_{c}(\theta)=\big\langle \prod_{i\in M}e^{i\hat{n}_{i}\theta} \big\rangle ,\ X_{s}(\theta)=\big\langle \prod_{i\in M}e^{i\hat{m}_{i}^{z}\theta} \big\rangle
\label{eq:disop_def}
\end{equation}
where $\hat{n}_{i}=\sum_{\sigma}\hat{c}_{i\sigma}^{\dagger}\hat{c}_{i\sigma}$ and $\hat{m}_{i}^{z}=\sum_{\sigma}\sigma\hat{c}_{i\sigma}^{\dagger}\hat{c}_{i\sigma}=2\hat{s}^z_i$
are the density  and  magnetization  along the $z$ quantization axis.   The operator products $\prod$ are performed in the region
$M$ shown in Fig.~\ref{fig:qsh-phase-diagram} (d) and (e). From the definitions in Eq.~\eqref{eq:disop_def}, it easily follows that 
$X_{c/s}(\theta)=X_{c/s}(\theta+2\pi)$, and $X_c(\pi)=X_s(\pi)$. A more detailed derivation of the disorder operator is presented in Sec.I of the Supplementary Materials (SM)~\cite{suppl}.  We  use   the  
ALF implementation~\cite{ALFSciPost_v2}  of the auxiliary field QMC algorithm to study the microscopic model of Eq.~\eqref{eq:model}   and  consider  linear system sizes $L=6,9,12,15,18$  at $\beta=L$.  

To further study the GN-QCP and DQCP and  for  comparisons, we consider  two other interacting models. For the  GN-QCP, we study  Dirac fermions based on the $\pi$-flux square lattice  with fermion-spin coupling model  that   triggers a  GN-Ising QCP towards a QSH phase~\cite{he2018Dynamical,liuDesigner2020,wangEmus2023}. The phase diagram and the model are shown in Fig.~\ref{fig:qsh-phase-diagram} (b) and (e), respectively.  The numerical results are shown in the Sec.III in SM~\cite{suppl}.

In addition, we consider a 1D DQCP spin model~\cite{huangEmergent2019,robertsDeconfined2019} with the Hamiltonian written as follows, $\hat H = \sum_i $

\noindent $\left(-J_x \hat S_i^x \hat S_{i+1}^x-J_z \hat S_i^z \hat S_{i+1}^z\right) + \left(K_x \hat S_i^x \hat S_{i+2}^x+K_z \hat S_i^z \hat S_{i+2}^z\right)$,

\noindent which describes a spin chain containing  nearest neighbor ferromagnetic interactions $J_x, J_z$ and  next-nearest neighbor antiferromagnetic interactions $K_x, K_z$. The model possesses a discrete $\mathbb{Z}_2^x \times \mathbb{Z}_2^z$ symmetry. We fix $K_x = K_z = 1/2$ and $J_x = 1$, such that the zero temperature phase diagram is described by only one parameter $J_z$, shown in Fig.~\ref{fig:qsh-phase-diagram} (c). The small/large $J_z$ limits are valence-bond-solid(VBS)/ferromagnetic(FM) phases, separated by a DQCP located at $J_c$. The emergent symmetry here is $[\mathrm{U}(1) \times \mathrm{U}(1)]\rtimes \mathbb{Z}_2$.  Previous infinite DMRG simulations find $J_c=1.4645$~\cite{huangEmergent2019,robertsDeconfined2019}, and we compute the disorder operator with DMRG to verify the presence of emergent continuous symmetry. 



\begin{figure}[htp!]
\includegraphics[width=\columnwidth]{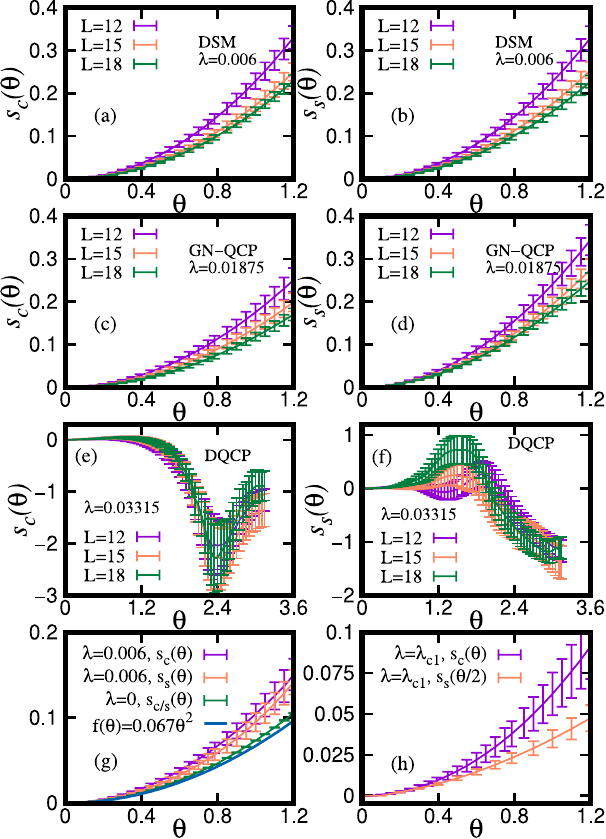}
\caption{Logarithmic coefficient $s_{c/s}(\theta)$ in the scaling of the disorder operator as a function of $\theta$ in DSM phase (a), (b), at the GN-QCP at $\lambda_{c1}$ (c), (d) and at the DQCP $\lambda_{c2}$ (e), (f). Different lines represent different system sizes $L$. The logarithmic coefficient $s_{c/s}(\theta)$ extrapolated to the thermodynamic limit are presented at (g) in the  DSM phase and (h) at GN-Heisenberg QCP.}
\label{fig:logcf_DSM-QSH}
\end{figure}

\noindent{\textcolor{blue}{\it DSM}---}
First, we present the results in the DSM phase of model in Eq.~\eqref{eq:model}.  From the measured $\left|X_{c}(\theta)\right|$ and $\left|X_{s}(\theta)\right|$ as a function of the perimeter $l$ (see Sec.II of SM~\cite{suppl}), one can fit the data with the scaling form of Eq. \eqref{Xscaling}, and extract the universal coefficients $s_{c/s}(\theta)$. The results are presented in Figs.~\ref{fig:logcf_DSM-QSH} (a)  and  (b).  
We note that when $\lambda=0$, there is an exact partial particle-hole symmetry $\hat{c}_\uparrow\rightarrow \hat{c}_\uparrow, \hat{c}_\downarrow \rightarrow \hat{c}_\downarrow^\dag$, under which $\hat{n}_i\rightarrow \hat{m}_i^z$. It is broken explicitly by the interaction term for $\lambda \neq 0$. Nevertheless, since the interaction is irrelevant in the DSM phase, the symmetry is still present in the IR theory (i.e. it is emergent), and therefore we expect that in  the IR $s_s(\theta)   = s_c(\theta) $  for  all values of the angles throughout the entire DSM phase.
The   result  of  the  size  extrapolation   $s_{s/c}(\theta) $    is  shown in  Fig.~\ref{fig:logcf_DSM-QSH}(g).    Within our  error bars  we have
$s_s(\theta)   = s_c(\theta) $   and  compare  well  with the  free case ($\lambda=0$).
It is known at small angle limit $\theta\rightarrow0$, $s_{c/s}(\theta)$ satisfies the quadratic form $s_{c/s}(\theta)=\alpha_{c/s}\theta^{2}$. The IR CFT gives $\alpha_{c/s} = \frac{A N_{\sigma} C_{J,\text{free}}}{8\pi^2}$, where $N_\sigma = 2$ is the spin flavor, $A \approx 1.30$ is a constant determined by the shape of the region~\cite{wu2021universal} (see SM~\cite{suppl} for more details on $A$). $C_J$ is the current central charge of the CFT.
     For the DSM phase we have $C_{J,\text{free}}=2$ as  for  free Dirac fermions~\cite{helmesUniversal2016}.
	 Extrapolation of  $\alpha_{c/s}$  is  shown  in Fig.~\ref{fig:alpha_vs_lambda}   and  we  obtain,  $\alpha_c(\infty)= 0.068(24)  $ and  $\alpha_s(\infty) = 0.068(31) $, fully consistent with the theoretical expectation $\alpha_{c/s} \approx 0.066$.

%

\noindent{\textcolor{blue}{\it QSH, SC }---}
The scaling behavior of the disorder operator in the QSH and SC phases is affected by the
gapless Goldstone modes originating from continuous symmetry breaking.   Our  data   for  $\left|X_{s}(\pi/2)\right|$ and $\left|X_{c}(\pi/2)\right|$   can be  found in  Sec.~II of the  SM~\cite{suppl}.    Given our  system  size  we  find it challenging  to  distinguish  between   additive,  $\ln l$,  and   
 multiplicative,  $l \ln l $, logarithmic corrections.  The latter  is  expected   for Goldstone modes as observed in the superfluid phase of Bose-Hubbard model~\cite{wangScaling2021}.



\begin{figure}[htp!]
\includegraphics[width=\columnwidth]{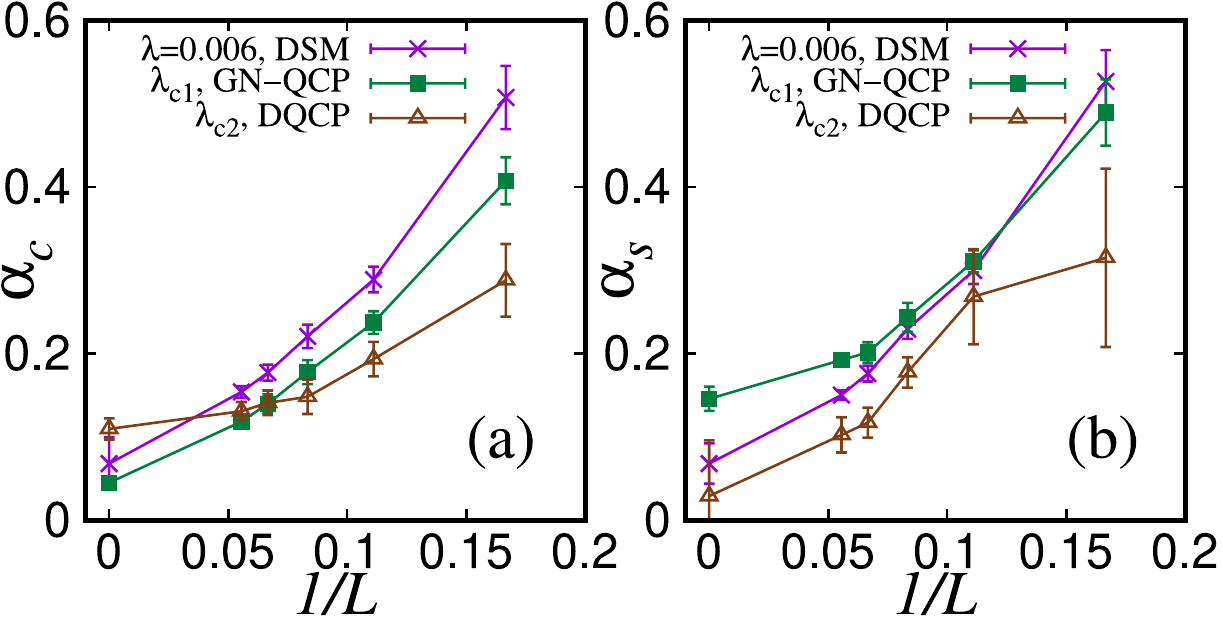}
\caption{
	System-size dependence of $\alpha_{c/s} = s_{c/s}(\theta)/\theta^{2}$, with $s_{c/s}(\theta)$ obtained from the fitting in Fig.~\ref{fig:logcf_DSM-QSH}. In the DSM phase ($\lambda=0.006$), extrapolation to the thermodynamic limit $L\to +\infty$ gives $\alpha_{c/s}=0.06(1)$, consistent with the expected CFT value of $\alpha = \frac{A N_{\sigma} C_{J,\text{free}}}{8\pi^2} \approx 0.066$.
	}
\label{fig:alpha_vs_lambda}
\end{figure}



\noindent{\textcolor{blue}{\it{DQCP  }---}
There  is  a  body  of  work  suggesting  emergent  SO(5)  symmetry   at  the  DQCP,   $\lambda_{c2}=0.03315$,  \cite{Nahum15_1, Nahum2015} 
and  it  is   intriguing  to    study  this  
from the point of  view  of  the  disorder operator.   To this aim, our  model is  {unusual since   the  U(1)  symmetry  is  embedded  as 
charge  conservation  and is  present at  the UV  scale (we note the same microscopic U(1) symmetry also occurs in a closely related cubic dimer model~\cite{sreejithEmergent2019}).}   
The  conjectured emergent  SO(5)  symmetry  at the DQCP  implies  that  
 the low  energy  theory  is invariant under  SO(5)   rotations  of the   super-spin vector,  
 consisting  of   the  two components of the superconducting order parameter  and  the  three  components   
 of  the  QSH  one  \cite{Tanaka05,Senthil06}.  
 Since  the  charge (spin)  disorder operator   of Eq.~\eqref{eq:disop_def}    rotates the superconducting   (QSH) 
  order  parameter  by   $2\theta$  one  expects   $s_{s}(\theta) = s_{c}(\theta)   $  for  all  values of  $\theta$.  
In  the  small  angle  limit  this    results in   $\alpha_s =  \alpha_c$  which  stands  at odds  with our  results:
  $\alpha_c(\infty) = 0.11(1) $ and   $\alpha_s = 0.022(63)$ albeit with very  large  error-bars. 
Furthermore,   at  larger  values of  the  angle,   $ s_{s/c} (\theta)$   go  negative  and  differ  substantially.      
Negative values of  $s_{c/s}(\theta)$   suggests  
 that the  DQCP  can not be described by a unitary  CFT,   it is a pseudo-critical point.    Similar  observations were  made in the JQ model, both in  the spin disorder operator~\cite{wang2021ScalingDQCP} and entanglement entropy $S_2$~\cite{JRZhao2022}. 

\noindent{\textcolor{blue}{\it{GN transitions }---}
In  contrast  to the   DQCP    $s_{c/s}(\theta) $  at  the  GN-Heisenberg transition  remains  positive    in finite  systems,  as shown in Fig.~\ref{fig:logcf_DSM-QSH}(c), (d),    and  after   extrapolation  to the    thermodynamic limit, shown in Fig.~\ref{fig:logcf_DSM-QSH} (h).  
This  confirms the  well-established picture  that  the GN-Heisenberg transition is described  by a  unitary  CFT.
For  the GN-Ising  transition discussed  in Sec. III of SM~\cite{suppl}   we  find  that the central charge $C_J$ 
is slightly reduced compared with that of the free DSM, consistent with the field-theoretical prediction~\cite{lliesiuBootstrapping2018}. 

At  the   GN-Heisenberg  critical point,    partial  particle-hole  symmetry  is  not present.    This  is    illustrated  by  the   fact  that  
$s_c(\theta)  \ne  s_s(\theta)$.      Interestingly,  our  data  seem to support  the  relation  $s_c(\theta)  \simeq  s_s(\theta/2)$.   

\begin{figure}[htp!]
\includegraphics[width=\columnwidth]{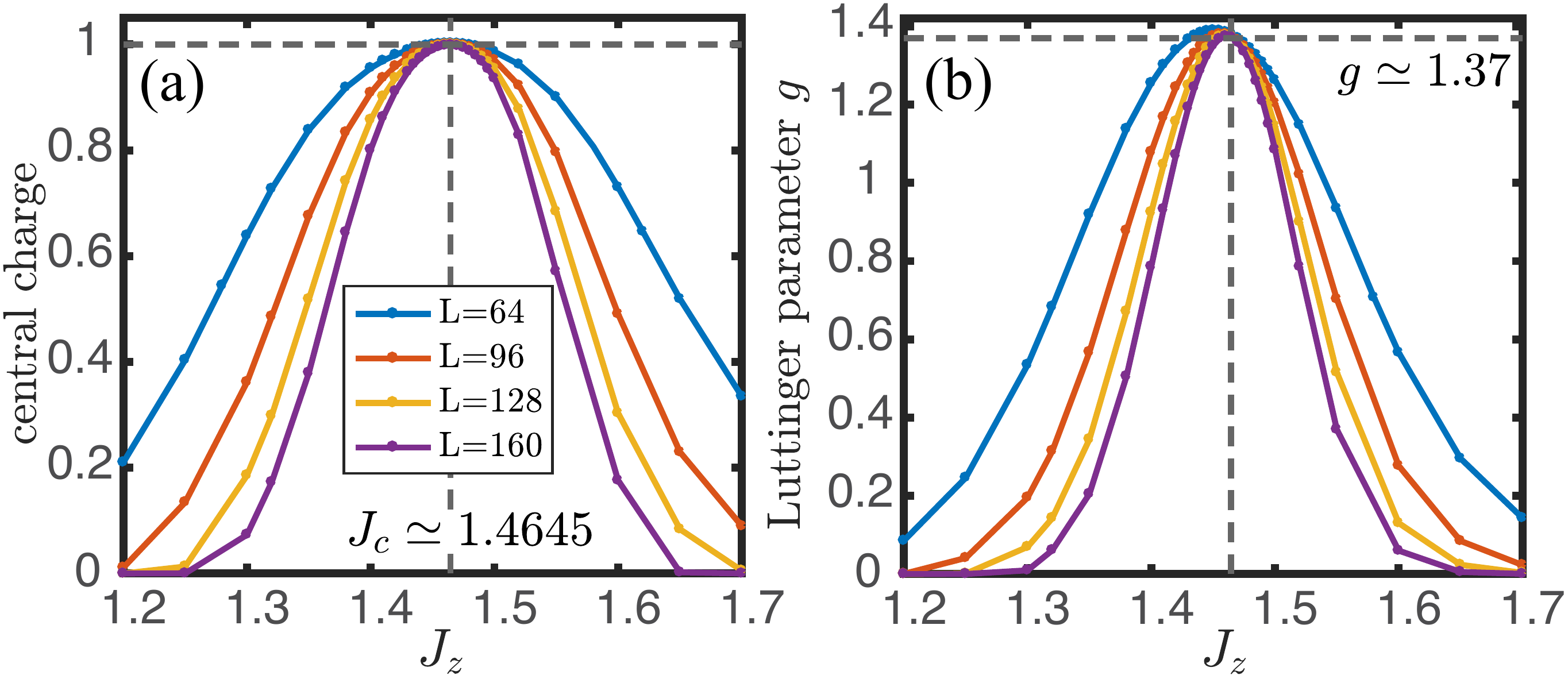}
\caption{
	Emergent symmetry at the 1D DQCP  $J_c=1.4645$. The (a) central charge $c$ and (b) Luttinger parameter $g$ are obtained from the CFT scaling behavior of entanglement entropy $S_\mathrm{vN}=\frac{c}{3}\ln{\tilde l}$ and disorder operator $-\ln{|X_s|}=\frac{g}{8}\ln{\tilde l}$, with the conformal distance $\tilde l= \frac{L}{\pi}\sin\frac{\pi l}{L}$, which indeed provide the correct values for $c$ and $g$ at the DQCP.	
 }
\label{fig:DQCP_1D}
\end{figure}

\noindent{\textcolor{blue}{\it  1D DQCP}---}
As  a  final  example,  we  study the 1D DQCP model~\cite{huangEmergent2019,robertsDeconfined2019} of Fig.~\ref{fig:qsh-phase-diagram} (c) with DMRG simulations.  Our  aim is to  further  confirm  that disorder operators follow the predictions of the IR theory with emergent  continuous  symmetries.  In Ref.~\cite{huangEmergent2019}, the continuous phase transition between ferromagnetic $z$-FM and VBS phases is determined with high precision. The DQCP point has emergent $[\mathrm{U}(1)\times\mathrm{U}(1)]\rtimes\mathbb{Z}_2$ symmetry with the $x$-FM and $y$-AFM constituting the first $\mathrm{U}(1)$, and $z$-FM and VBS the second. The Luttinger parameter is determined in previous work at the DQCP from two-point correlation functions to be $g=1.38(1) $~\cite{huangEmergent2019}. Since the lattice model only has discrete symmetries, we compute the spin disorder operator defined as $X_s=\langle \prod_{i\in M} \sigma^{z}_i \rangle$ (corresponding to $\theta=\pi$) for a region $M$ of length $l$, and extract the scaling behavior from $-\ln (|X_{s}(l)|)= \frac{g}{8} \ln l$, as  has  been done for the  1d Hubbard model~\cite{jiangFermion2022}~\footnote{In our DMRG calculations, we take the periodic boundary condition and confirm the data (including von Neumann EE and disorder operator) are well converged with up to $D=1024$ bond states kept in the simulations}. 
In Fig.~\ref{fig:DQCP_1D}(a), the central charge $c=1$ around the DQCP point $J_z\simeq1.4645$, 
is extracted from $S_\mathrm{vN}=\frac{c}{3}\ln{\tilde l}$ 
with the conformal distance $\tilde l = \frac{L}{\pi}\sin\frac{\pi l}{L}$ and in Fig.~\ref{fig:DQCP_1D} (b), the Luttinger parameter $g\simeq1.37$ is extracted from the disorder operator on the finite-size systems, following the finite-size scaling form $-\ln(|X_s(l)|)=\frac{g}{8}\ln \tilde l$. 
Away  from the  critical point,   and as supported  by the data,  the log  corrections  are    expected  to  vanish  since all correlation functions  are  characterized by a finite length  scale.  


\noindent{\textcolor{blue}{\it Discussion.}---}
The  disorder operator is  a  simple  quantity to implement in many  numerical  approaches.   Especially  in the 
realm of   DQMC and in comparison  to  EE and ES    the  disorder operator is  very  easy  to access  and  can 
be  computed \textit{on the fly}.    While,  1D and 2D boson/spin systems 
as  well as nearly free fermion systems 
have been intensively investigated from the point of  view of entanglement, ~\cite{helmesEntanglement2014,vidalEntanglement2003,poilblanc2010entanglement,hastingsMeasuring2010,pollmannEntangle2010,fidkowski2010Entanglement,dEmidio2020Entanglement,humeniuk2012Quantum,inglis2013Wang,luitz2014Improving,inglisNJP2013Entanglement,kallinJS2014Corner,kallinPRL2013Entanglement,Fradkin2006,Casini2006,Kitaev2006top,Levin2006,liEntanglement2008,poilblanc2010entanglement,dEmidio2020Entanglement,JRZhao2022,yanRelating2021,JRZhao2022QiuKu,demidioUniversal2022,songReversing2022},  the simplest interacting fermion lattice models in 2D, that is,  Dirac fermions and their interaction-driven Gross-Neveu (GN) transitions~\cite{chandrasekharan2013Quantum,otsukaUniversal2016,zerf2017Four,Ihrig2018Critical,he2018Dynamical,langQuantum2019,liu2019superconductivity,liuDesigner2020,schulerTorus2021,liuGross2021,tabatabaeiChiral2022,liuMetallic2022,erramilliGross2022,liaoDiracI2022,liaoDiracII2022,liaoDiracIII2022,wangEmus2023} remain   
challenging.   This is mainly due to the fact that, the EE, for example the $n$-th order R\'enyi entropy $S_n$, needs to be computed in path-integral QMC simulations with a replicated manifold. 
Since the typical DQMC simulation for interacting fermions is already very expensive (usually the computational complexity scales as $O(\beta N^3)$ with $\beta=1/T$ and $N=L^D$) 
 the construction of the replicated manifold and the ensemble average therein~\cite{Assaad2014,Assaad2015,Parisen2018}  
 is  challenging. 
As a result, small  system sizes in 2D and often noisy data  is insufficient to extract  universal scaling properties  in EE. We note  recent  progress in this regard, with better data quality and the approximate $O(\beta N^3)$ scaling~\cite{demidioUniversal2022,liaoTeaching2023,panComputing2023}. 
   
In  contrast  no  replica  construction  is   required for the   disorder operator.   
Logarithmic  corrections to the  area  law  capture  IR  physics  and provide  invaluable insights  into critical points.  
Although it  is  challenging  to  control   logarithmic  corrections  to  the area law,  especially  in  DQMC,   
 we  have  presented  a large body of   data  that  shows consistent  results  for  Dirac  systems.    Particular   we  can test for  emergent    symmetries  at  the  DSM point  by  comparing 
 particle-hole  related   disorder operators.    One  can also  extract  the central charge $C_J$.    Emergent symmetries  at  a  1D  DQCP 
 can  equally be  tested.     The  CFT constraint on the  sign of the  coefficient of the logarithmic correction,   $s(\theta) > 0$,   allows  to  
 test  if  a   putative  critical point  actually  corresponds to a  unitary  CFT.    Using this  criterion we  are  able to  show  
 that  our  realization of  a  monopole  free DQCP    does  not seem to   correspond to  a  unitary CFT.  This is  consistent  with  results  for the 
 JQ model~\cite{wang2021ScalingDQCP,JRZhao2022}    and  hints  to the power  of  the     disorder  operator. 
 Given  the  \textit{simplicity}   in computing  this quantity,  especially within the realm of  fermion QMC,  many new directions are opened from here, concerning  the enigmatic fate of the 2D DQCP theories and their lattice model realizations~\cite{liaoDiracI2022,liaoDiracII2022,sandvikEvidence2007,maRole2019,kuklovDeconfined2008,chenDeconfined2013,emidioDiagnosing2021,nahumNote2020,maTheory2020,gorbenkoWalking2018,zhaoMulticritical2020,liaoTeaching2023}.

{\it{Acknowledgment.-}} WLJ, BBC and ZYM acknowledge the support from the Research Grants Council of Hong Kong SAR of China (Project Nos. 17301420, 17301721, AoE/P-701/20, 17309822, HKU C7037-22G), the ANR/RGC Joint Research Scheme sponsored by Research Grants Council of Hong Kong SAR of China (Project No. A\_HKU703/22) and French National Research Agency (grant ANR-22-CE30-0042-01). The Strategic Priority Research Program of the Chinese Academy of Sciences (Grant No.
XDB33000000), the K. C. Wong Education Foundation (Grant No.~GJTD-2020-01). FFA  and  ZL   acknowledge financial support from the DFG through the W\"urzburg-Dresden Cluster of Excellence on Complexity and Topology in Quantum
Matter - \textit{ct.qmat} (EXC 2147, Project No.\ 390858490)   as  well as  the SFB 1170 on Topological and Correlated Electronics at Surfaces and Interfaces (Project No.\  258499086).
We thank the HPC2021 platform under the Information Technology Services at the University of Hong Kong, and the Tianhe-II platform at the National Supercomputer Center in Guangzhou for their technical support and generous allocation of CPU time.  We equally  gratefully acknowledge the Gauss Centre for Supercomputing e.V. (www.gauss-centre.eu) for funding this project by providing computing time on the GCS Supercomputer SuperMUC-NG at Leibniz Supercomputing Centre (www.lrz.de).

\bibliographystyle{shortapsrev4-2}
\bibliography{qsh}


\clearpage

%


%
%
%
%
%

\setcounter{figure}{0}
\setcounter{equation}{0}
\renewcommand\thefigure{S\arabic{figure}}
\renewcommand\theequation{S\arabic{equation}}

\title{Supplemental Material for \\[0.5em]
``Fermion disorder operator at Gross-Neveu and deconfined quantum criticalities''}

\begin{abstract}
In this supplemental material, we provide details of simulations, the original data, related theoretical analysis to offer helpful context to coordinate with the main text. The supplemental material is organized as follows:  In Section  \ref{sec:qmc},  we provide the details of the QMC simulations and properties on the disorder operator. In Sections \ref{sec:2ddqcp}, \ref{sec:secVI} and \ref{sec:1ddqcp}, we will discuss 2d DQCP model, the $\pi$-flux model, and  1d DQCP model  respectively. 
\end{abstract}

\maketitle

\section{QMC simulations and the disorder operator}
\label{sec:qmc}

\subsection{Quantum Monte Carlo implementation}
In this work, we use the ALF~\cite{ALFSciPost_v2} implementation of DQMC at finite-temperature. We used a symmetric Suzuki-Trotter decomposition to control the systematic error in observables and adopted an imaginary time step $\Delta\tau t=0.2$. 
 We set the parameter $t=1$ and scale the inverse of temperature
$\beta$ as $\beta=L$ in the calculation to access the thermodynamic
limit. The coupling strength $\lambda$ is the parameter we tune in calculation. The disorder operators are defined in parallelogram
region in honeycomb lattice as shown in Fig. 1 (c) in the main text.

The measurement of  the disorder operator in DQMC  can be  implemented as \cite{Assaad08_rev}
\begin{align}
X\left(\theta\right)  & =\left\langle \prod_{\boldsymbol{i}\in M}\exp\left(i\hat{Q}_{\boldsymbol{i}}\theta\right)\right\rangle =\frac{1}{Z}\text{Tr}\left\{ e^{-\beta H}e^{i\hat{c}^{\dagger}T\left(\theta\right)\hat{c}}\right\} \nonumber \\
 & =\sum_{\left\{ s\right\} }P_{s}\det\left(\boldsymbol{1}+\Delta\left(\theta\right)\left(\boldsymbol{1}-G_{M,s}\right)\right)
\label{eq:disop_cal_qmc}
\end{align}
for the fermion bilinear local operator $\hat{Q}_{\boldsymbol{i}}$, where $G_{M,s}$ is the block Green's function matrix in subregion $M$  for a given configuration of  Hubbard-Stratonovich fields, $s$.  Since  the  Green  function matrix  is   at  hand,   the   disorder operator  can be computed  on the 
fly  without   enhancing  the  computational  effort.   The matrix elements of $\Delta\left(\theta\right)=e^{iT\left(\theta\right)}-\boldsymbol{1}$ dependent on the form of local operator. In our calculation, we consider the local operator to be $\hat{Q}_{\boldsymbol{i}}=\hat{n}_{\boldsymbol{i}}$ and $\hat{Q}_{\boldsymbol{i}}=\hat{m}_{\boldsymbol{i}}$. If the Hamiltonian we consider is block diagonal in spin basis, we can write down the expression Eq.(\ref{eq:disop_cal_qmc}) as the product of the determinant, 
\begin{equation}
X_{c/s}\left(\theta\right) =\sum_{\left\{ s\right\} }P_{s}\prod_{\sigma=\uparrow\downarrow}\det\left(\boldsymbol{1}+\Delta_{\sigma}^{c/s}\left(\theta\right)\left(\boldsymbol{1}-G_{M,s,\sigma}\right)\right)
\end{equation}
where $\left(\Delta_{\sigma}^{c}\right)_{\boldsymbol{ii}^{\prime}\in M}\left(\theta\right)=\delta_{\boldsymbol{ii}^{\prime}}\left(e^{i\theta}-1\right)$ and $\left(\Delta_{\sigma}^{s}\right)_{\boldsymbol{ii}^{\prime}\in M}\left(\theta\right)=\delta_{\boldsymbol{ii}^{\prime}}\left(e^{i\sigma\theta}-1\right)$.
For the non-interacting system, $G_{M,s}$ is not dependent  on the configuration of  Hubbard-Stratonovich fields  and the Eq.~(\ref{eq:disop_cal_qmc}) reduces to Eq.(\ref{eq:cal_mft_disop}). 

\subsection{Partial particle-hole symmetry}
In this part we provide a mapping between the disorder operators $\left| X_{c}(\theta)\right|$ and $\left| X_{s}(\theta)\right|$ under partial particle-hole symmetry. We first introduce the definition of the particle-hole symmetry as $P$: $\hat{c}_{\boldsymbol{i}\uparrow}\rightarrow \hat{c}_{\boldsymbol{i}\uparrow}$, $\hat{c}_{\boldsymbol{i}\downarrow}\rightarrow (-1)^{\boldsymbol{i}}\hat{c}^{\dagger}_{\boldsymbol{i}\downarrow}$. One can easily obtain the follow relation
\begin{align}
P\hat{m}_{\boldsymbol{i}}P^{-1} & =P\left(\hat{c}_{\boldsymbol{i}\uparrow}^{\dagger}\hat{c}_{\boldsymbol{i}\uparrow}-\hat{c}_{\boldsymbol{i}\downarrow}^{\dagger}\hat{c}_{\boldsymbol{i}\downarrow}\right)P^{-1}\nonumber \\
 & =\hat{c}_{\boldsymbol{i}\uparrow}^{\dagger}\hat{c}_{\boldsymbol{i}\uparrow}-\hat{c}_{\boldsymbol{i}\downarrow}\hat{c}_{\boldsymbol{i}\downarrow}^{\dagger}=\hat{n}_{\boldsymbol{i}}-1
\end{align}
We further consider a model hamiltonian with partial particle-hole symmetry such that $P\hat{H}_0P^{-1}=\hat{H}_0$. Combining with the above equation we have the relation $ X_{s}\left(\theta\right) =e^{-iN_{s}\theta} X_{c}\left(\theta\right)$, where $N_s$ represent the total number of site inside the region M.  Since the measurement of the disorder operators are based on the absolute values $\left| X_{c/s}(\theta)\right|$, the additional phase factor generated by transformation $P$ does not change the result.  In summary, the disorder operators  $\left| X_{c}(\theta)\right|$ and $\left|X_{s}(\theta)\right|$ are the same in the model with partial particle-hole symmetry.

\section{2d DQCP model}
\label{sec:2ddqcp}
First, we focus on the model whose Hamiltonian is introduced in Eq.(1) in the main text. To begin, the  mean field limit of the Hamiltonian is discussed. Next, we display our DQMC results. We use the RG invariant quantities to determine $\lambda_{c1}$ and $\lambda_{c2}$, consistent with the results of previous papers. The unprocessed simulation results on the disorder operator and fitting results for $\alpha$ for the  interacting case are  then provided.

\subsection{Mean field limit}
\label{sec:MFT}

Besides the discussion of the QMC implementation of the disorder operator in the interacting model, in this section, we also provide a simple test by using mean field hamiltonians.   For  the honeycomb lattice we  induced  the orders in the phase diagram  by including  mass terms: 
\begin{align}
H_{\text{MF}}= & -t\sum_{\left\langle ij\right\rangle \sigma}\left(\hat{c}_{i\sigma}^{\dagger}\hat{c}_{j\sigma}+h.c.\right)+m\sum_{i\sigma}(-1)^{\boldsymbol{i}}\hat{c}_{i\sigma}^{\dagger}\hat{c}_{i\sigma}-\nonumber \\
 & \lambda\sum_{\hexagon}\boldsymbol{N}\cdot\left(\sum_{\left\langle \left\langle ij\right\rangle \right\rangle \in\hexagon}i\nu_{ij}\hat{c}_{i}^{\dagger}\boldsymbol{\sigma}\hat{c}_{j}+h.c.\right).
\label{eq:ham_mft}
\end{align}
Here,  $m$ is the staggered mass in real space that  generates the charge density wave (CDW) long range order. The mass term of  amplitude $\lambda$ is given by the vector product of the  O(3) vector $\boldsymbol{N}$ and the generalized spin-orbit coupling term, which produces the long-range QSH order. The  Hamiltonian  of Eq.~\eqref{eq:ham_mft} only contains fermion bilinear terms and can be solved exactly. For a given fermion bilinear Hamiltonian $H_{\text{bilinear}}=\hat{c}^{\dagger}K\hat{c}$, the disorder operator can be calculate  using  the following simple formula
\begin{align}
X_{\alpha}(\theta)  & =\frac{1}{Z}\text{Tr}\left\{ e^{-\beta\hat{c}^{\dagger}K\hat{c}}e^{i\hat{c}^{\dagger}T(\theta)\hat{c}}\right\}\nonumber \\
 & =\det\left(G+e^{iT(\theta)}\left(\boldsymbol{1}-G\right)\right)
 \label{eq:cal_mft_disop}
\end{align}
where the matrix $T(\theta)$ is a diagonal matrix with nonzero diagonal element when the matrix index  belong to the region $M$ and $G=\left(\boldsymbol{1}+e^{-\beta K}\right)^{-1}$ is the Green's function.  This  allows for an efficient calculation of  the determinant in Eq. (\ref{eq:cal_mft_disop}). In the following, we focus on the disorder operator $X_{c/s}(\theta)$ as defined in the main text and discuss their behavior for different mean field    Hamiltonians. 

\begin{figure}[tb]
	\begin{centering}
	\includegraphics[width=0.48\textwidth]{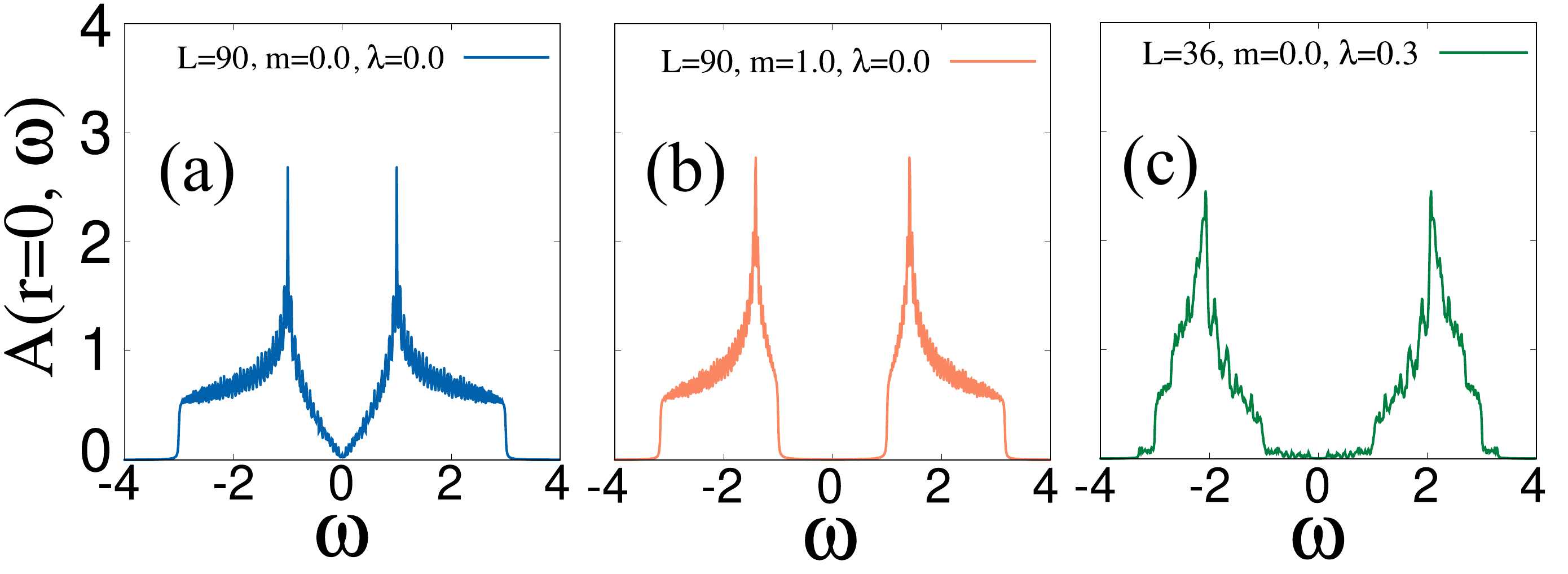}
	\par\end{centering}
	\caption{Local density of state(LDOS) of  the model Eq.~(\ref{eq:ham_mft}) with different model parameters. (a) is the free fermion LDOS for the honeycomb lattice with system size $L=90$. In (b) we consider finite the charge density wave mass by setting $m=1$ and $\lambda=0$. In (c) we consider the finite QSH type mass by setting $\lambda=0.3$ and $m=0$.}
	\label{fig:mft_dos}
	\end{figure}
	
Fig.~\ref{fig:mft_dos} depicts the  local density of states (LDOS) of  the model in Eq.~\eqref{eq:ham_mft} at  various   mean-field parameters. By setting $t=1$, $m=0$ and $\lambda=0$, Eq.~\eqref{eq:ham_mft} describe a stable Dirac semi-metal ground state in the  thermodynamic limit, as presented in Fig.~\ref{fig:mft_dos} (a). Without the mass term, the Hamiltonian is block diagonal in the  spin basis so that the disorder operator in the  charge  and spin channels  are identical.  We calculate the disorder operator with the parallelogram region as in the Fig.1 (c) in the main text. In Fig.~\ref{fig:mft_DSM_disop} (a), for a given rotation angle $\theta$,  the disorder operator $X(\theta)$ is  dominated by the perimeter law decay, $\text{ln}\left|X_{c/s}\left(\theta\right)\right|\sim -al+s(\theta)\ln l+c$. As we discussed above, the free Dirac system is a typical conformal fixed point and the disorder operator has  sub-leading logarithmic corrections~\cite{lliesiuBootstrapping2018,helmesUniversal2016}. We fit the data, Fig.~\ref{fig:fit_window_mft1},  to extract the 
 logarithmic coefficient $s(\theta)$ as a function  of angle,   Fig.~\ref{fig:mft_DSM_disop} (b). Apparently, at small angle, $s(\theta)\sim\alpha\theta^2$. At large system size, the quadratic coefficient $\alpha$ converges to a stable value $\alpha=0.067(5)$. This is comparable  to  $\alpha=0.0658$ obtained in the thermodynamic limit. It is  equally  consistent with the other free Dirac fermion computation for the $\pi$-flux model in Sec.~\ref{sec:secVI}. Specifically, we obtain the analytic density correlation function combining the corner distribution of the region $M$, where the detailed analysis is presented in Sec.~\ref{sec:secVII}. The  slight differences in our  results  stem  from finite size effect,  which can be eliminated by  extrapolation.  
	
\begin{figure}[tb]
	\begin{centering}
	\includegraphics[width=0.48\textwidth]{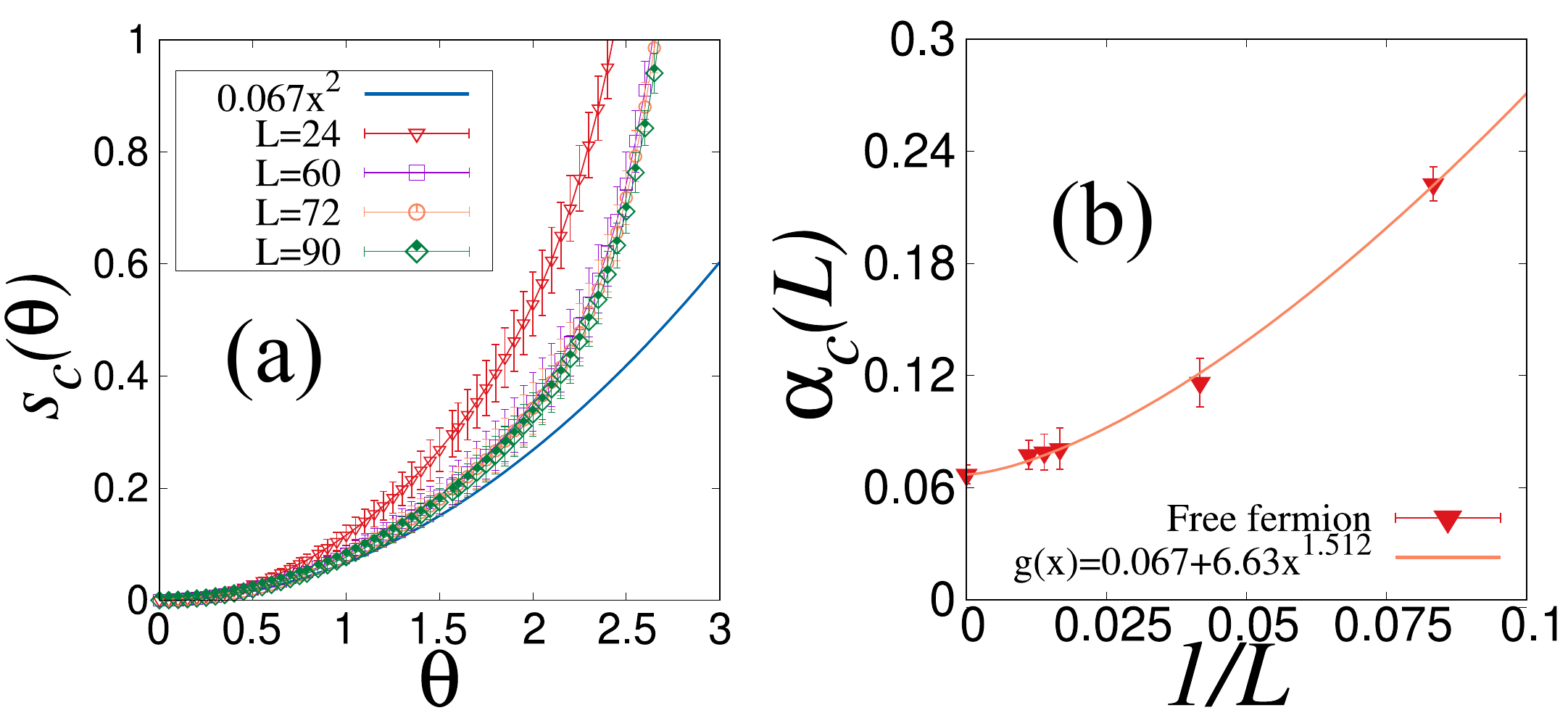}
	\par\end{centering}
	\caption{(a)Logarithmic correction term $s_c(\theta)$ extract from the disorder operator $X_c(\theta)$ by using the numerical fitting as function of rotation angle $\theta$. (b)The quadratic coefficient $\alpha(L)$ as function of $1/L$ from free fermion model.}
	\label{fig:mft_DSM_disop}
\end{figure}
	
\begin{figure}[tb]
\begin{centering}
\includegraphics[width=0.48\textwidth]{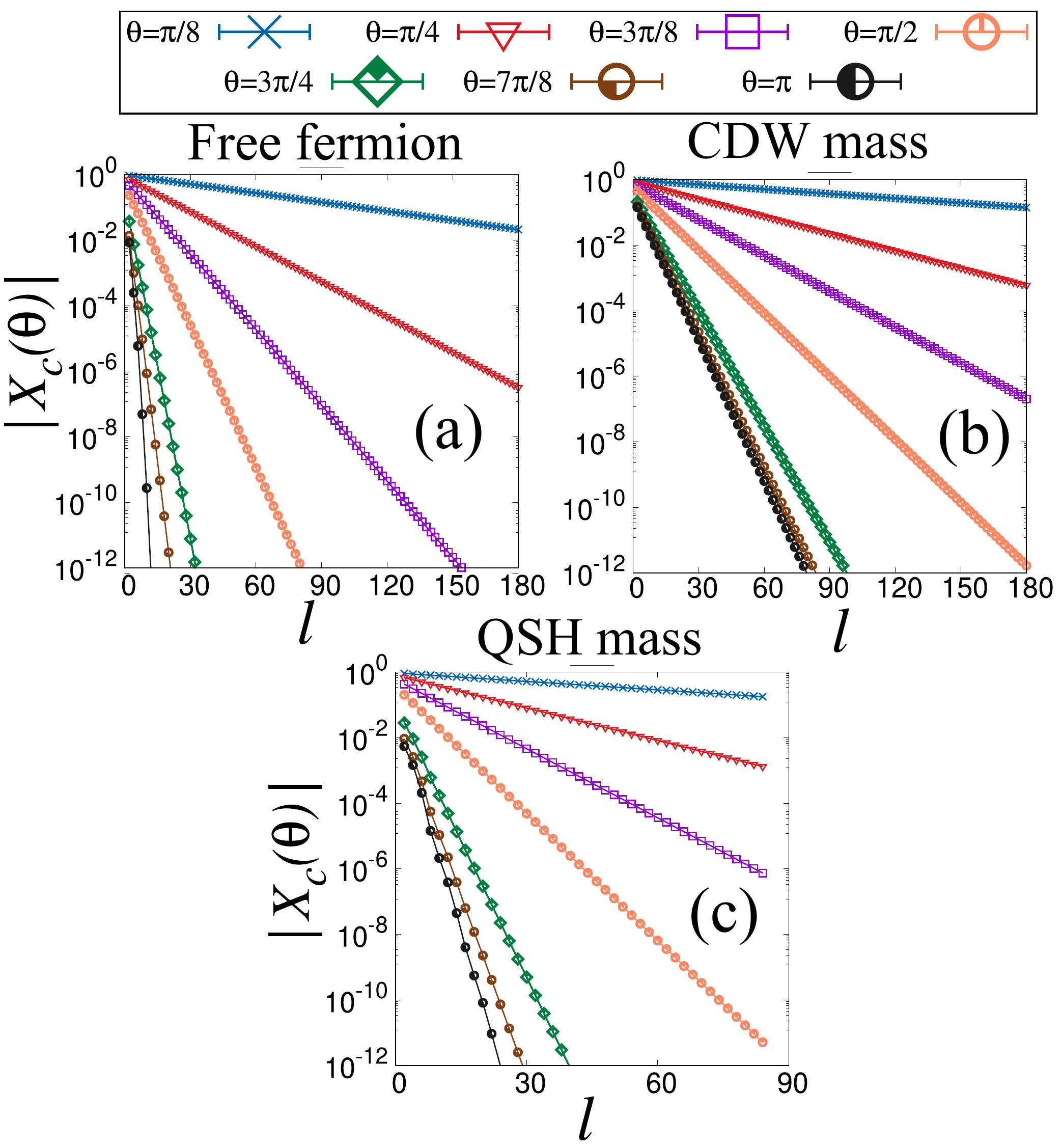}
\par\end{centering}
\caption{Disorder operator $X_c(\theta)$ of model Eq.(\ref{eq:ham_mft}) defined in charge channel as function of perimeter $l$. We set the y axis in logarithmic scale. In (a) we consider free fermion model in honeycomb lattice by setting $m=\lambda=0$. In (b) we consider the model with finite CDW mass $m=1.0$. In (c) we set $\lambda=0.3$ to observe the effect of QSH type mass term. We choice system size $L=90$ at (a)(b) and $L=36$ at (c).}
\label{fig:mft_mass_disop}
\end{figure}
	
In Fig.~\ref{fig:mft_mass_disop}, we present the disorder operator results for massive Dirac system. In Fig.~\ref{fig:mft_mass_disop} (b), we set  $t=1$, $m=1.0$ and $\lambda=0$, which turns on the staggered CDW mass term that breaks  sublattice symmetry. The CDW mass term is diagonal in the spin basis,  such  that  the  disorder operators  are channel independent. The disorder operator in  the massive fermion  exhibit  a  much slower decay  than the  gapless Dirac fermions, but is still dominated by  a perimeter law decay. In Fig.~\ref{fig:mft_mass_disop}(c), we set the parameter $t=1$, $m=0$, $|\lambda\boldsymbol{N}|=1$ and $N_x=N_y=N_z$ to turn on the QSH mass term. The QSH mass term breaks the SU(2) symmetry of free Dirac fermion which removes the degeneracy of two channels in the disorder operator. Due to the fact that the SU(2) symmetry break into U(1) symmetry correspond to the rotation around axis $\boldsymbol{N}$ in  the mean field  Hamiltonian, 
the spin disorder operator $X_s(\theta)$ is not well defined when $\boldsymbol{N}/|\boldsymbol{N}|\ne \boldsymbol{e}_z$.
 However,  the global charge conservation is  present  for  the  QSH mass such  that  $X_c(\theta)$ is still well defined.
  In Fig.~\ref{fig:mft_mass_disop}(c), the decay rate of $X_c(\theta)$ is dominated by  the perimeter law.     As  apparent from  
  Fig.~\ref{fig:fit_window_mft1}(c)  logarithmic corrections  to the area  law  are absent in the massive  CDW  phase. 
    
\begin{figure}[tb]
\begin{centering}
\includegraphics[width=0.48\textwidth]{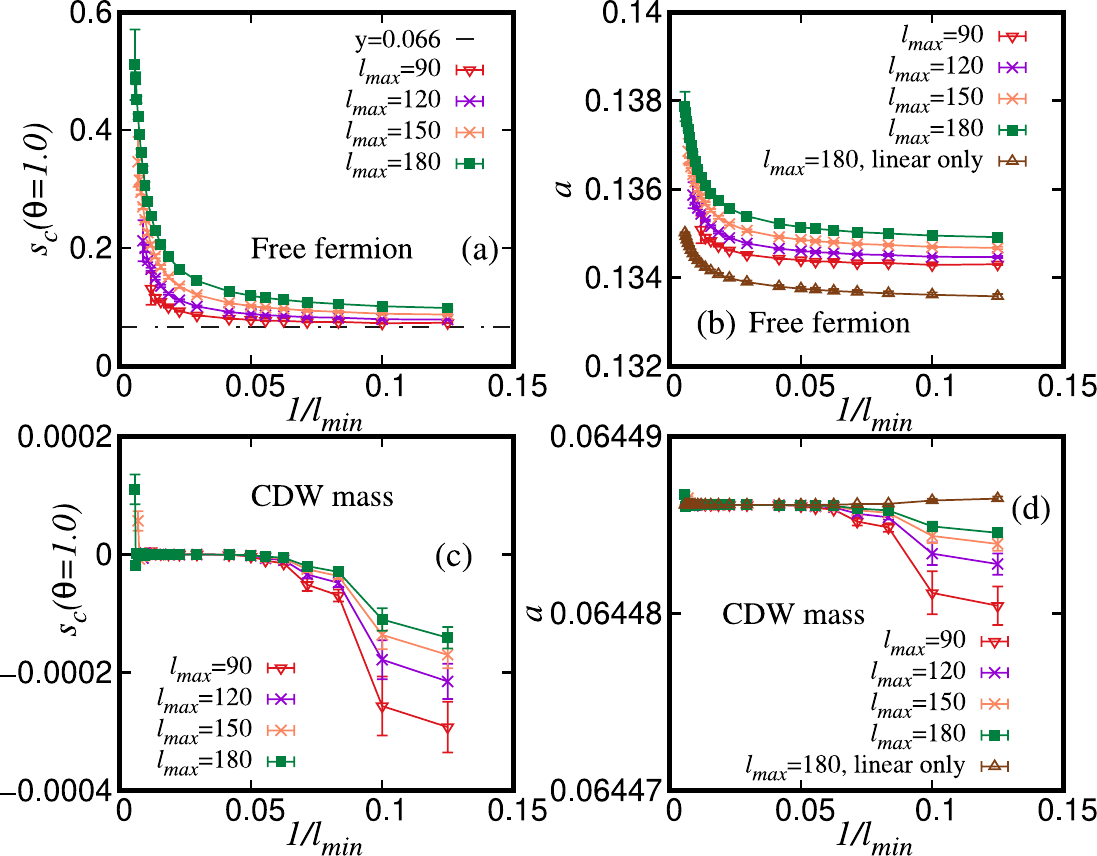}
\par\end{centering}
\caption{
The fitting  range  is  set by  $\left[ l_{min},  l_{max}  \right] $  and  the  data is     fitted  to the  form   
$  \ln \left|X(\theta)\right|  =  -al+s(\theta)\ln l+c. $
Logarithmic coefficient $s(\theta)$(a) (c) and linear coefficient $a$ (b) (d) as function of  the smallest perimeter $l_{min}$ considered in the  fitting.  
We set $L=90$ and $\theta=1.0$. (a) and (b) are obtained from  the free fermion model and (c)  and (d) are obtained from model with CDW mass term. The brown dot lines in (b) and (d) indicate the linear coefficient $a$ extract from the fitting of linear function.}
\label{fig:fit_window_mft1}
\end{figure}

\subsection{RG invariant quantities}
As a benchmark to previous calculations~\cite{liu2019superconductivity,liuGross2021}, we present the RG invariant  correlation ratio, 
$R_{c}^{\text{QSH/SC}}$ as a function of coupling constant $\lambda$ in Fig.\ref{fig:Rc_qsh}.  It  is defined as 
\begin{equation}
R_{c}^{\text{QSH/SC}}=1-\frac{S_{\text{QSH/U(1)}}\left(\boldsymbol{k}=\Gamma+d\boldsymbol{k},\tau=0\right)}{S_{\text{QSH/U(1)}}\left(\boldsymbol{k}=\Gamma,\tau=0\right)}
\end{equation}
where $d\boldsymbol{k}=(0,\frac{4\pi}{\sqrt{3}L})$ and $S_{\text{QSH/U(1)}}=\frac{1}{L^2}\sum_{\boldsymbol{r}\boldsymbol{r^{\prime}}}e^{i\boldsymbol{q}\cdot(\boldsymbol{r}-\boldsymbol{r}^{\prime})}\mathrm{Tr}\{\hat{O}_{\boldsymbol{r}}\hat{O}_{\boldsymbol{r}^{\prime}}\}$ is the structure factor defined by the order parameter $\hat{O}_{\boldsymbol{r}}$. The QSH local vector order parameter takes the form of a spin current $\hat{O}^{\text{QSH}}_{\boldsymbol{r},\boldsymbol{\delta}}=i\hat{\boldsymbol{c}}^{\dagger}_{\boldsymbol{r}}\boldsymbol{\sigma}\hat{\boldsymbol{c}}_{\boldsymbol{r}+\boldsymbol{\delta}}+\text{H.c.}$ where $\boldsymbol{\delta}$ runs over all the  next-nearest neighbours of the hexagon labeled by $\ve{r}$. We use the SC local order parameter $\hat{O}_{\boldsymbol{r},\tilde{\boldsymbol{\delta}}}^{\text{SC}}=\frac{1}{2}\left(\hat{c}_{\boldsymbol{r}+\tilde{\boldsymbol{\delta}},\uparrow}^{\dagger}\hat{c}_{\boldsymbol{r}+\tilde{\boldsymbol{\delta}},\downarrow}^{\dagger}+\text{H.c.}\right)$ where $\tilde{\boldsymbol{\delta}}$ runs over the sublattice a,b in the honeycomb lattice.

\label{sec:RG}
\begin{figure}[htb]
\begin{centering}
\includegraphics[width=0.48\textwidth]{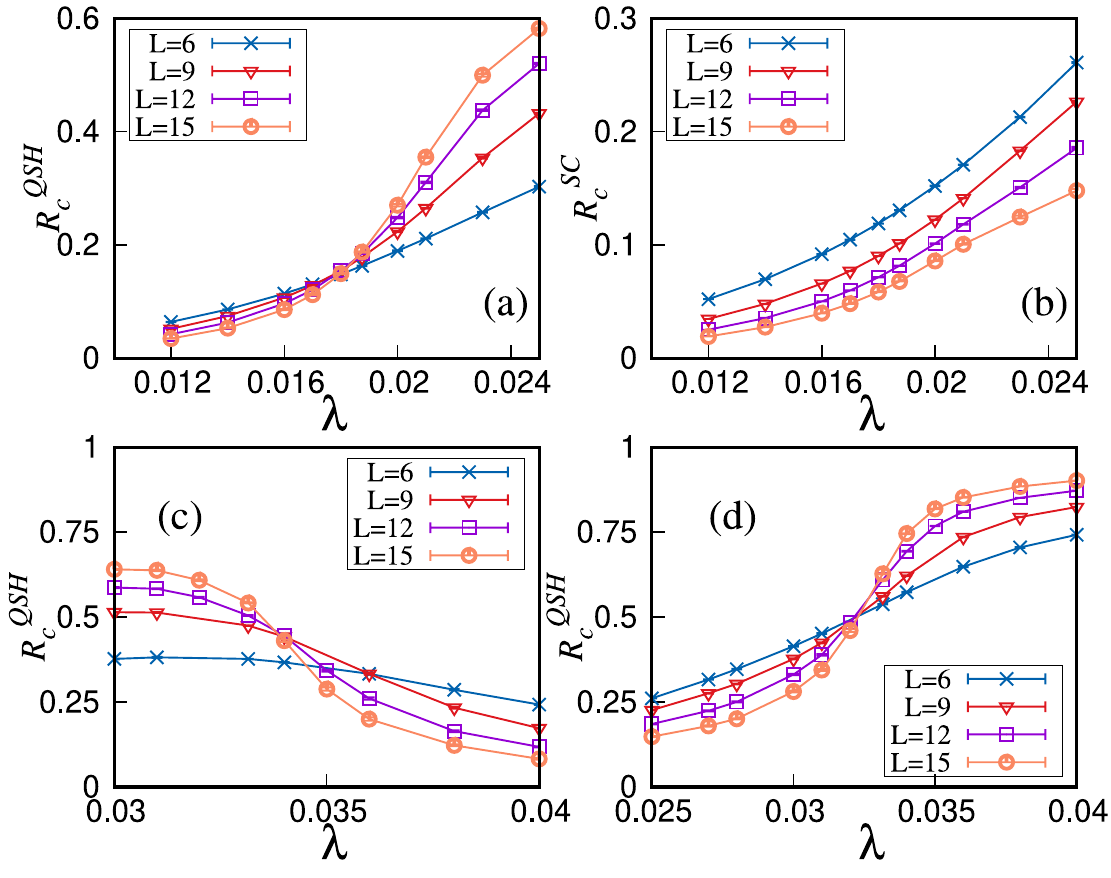}
\par\end{centering}
\caption{RG invariant quantities $R_{c}^{\text{QSH/SC}}$ as function of $\lambda$ for the  model of Eq.~(1) in the main text.  Due to the  expected  
Lorentz  symmetry  of  the  critical points,  we scale  the 
inverse temperature  with system size:  $\beta=1/T=L$.}
\label{fig:Rc_qsh}
\end{figure}

The results for $R_{c}^{\text{QSH}}$ in the vicinity of  the GN-QCP are presented in Fig.~\ref{fig:Rc_qsh}(a). The crossing point of $R_{c}^{\text{QSH}}$ is stable and  yields  $\lambda_{c1} = 0.0187(2)$. In Fig.~\ref{fig:Rc_qsh}(b), we observe that  $R_{c}^{\text{SC}}$ vanishes  upon  increasing 
system size in the vicinity of the  GN-QCP. This  reflects  the absence of the long-range s-wave pairing   at  this    critical  point. In Figs.~\ref{fig:Rc_qsh}(c) and (d), our QMC data suggest that the trends of the crossing points of the two RG invariant quantities $R_{c}^{\text{QSH}}$ and $R_{c}^{\text{SC}}$ 
converge to the  same  value  as a  function of  system size.  Again, our data is consistent with the estimate $\lambda_{c2} = 0.0332(2)$.

\subsection{Raw data of disorder operator in the interacting case}
\label{sec:Raw data}
In this subsection we show the behavior of the raw data for the  disorder operators $X_{c/s}(\theta)$ for  various coupling constants $\lambda$ and rotation angle $\theta$. In Fig.~\ref{fig:x-vs-theta}, the disorder operator $X_{c/s}(\theta)$ at fixed angle $\theta=\pi/8,\pi/4,\pi/2,3\pi/4,\pi$ illustrate the evolution as function of coupling $\lambda$. Up to  $L=15$, the decay rate of $X_{c/s}(\theta)$ is dominated by the area law $\sim \exp(-aL)$. As we increase the coupling, we observe a clear slowing down of the decay rate.

\begin{figure}[tb]
\begin{centering}
\includegraphics[width=0.46\textwidth]{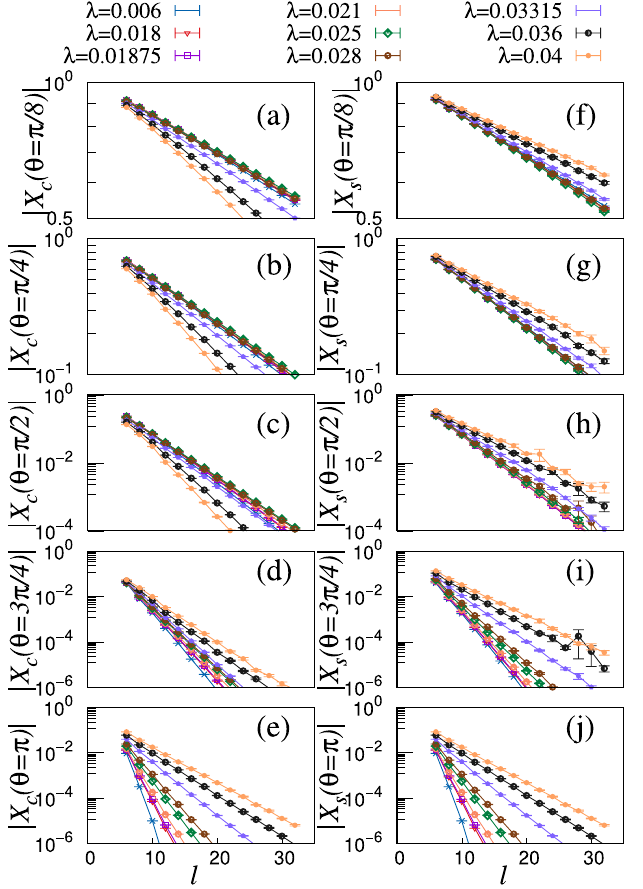}
\par\end{centering}
\caption{Disorder operator $X_{c}(\theta)$(a)-(e) and $X_s(\theta)$(f)-(j) obtained from  the model of  Eq.~(1) in the main text as function of 
perimeter $l$ on the  honeycomb lattice.  We consider a  logarithmic  y-axis  and  $L=\beta=15$. Different subplots correspond to different 
rotation angles $\theta$. In each subplot, we present the disorder operator obtained for different $\lambda$ for comparison. }
\label{fig:x-vs-theta}
\end{figure}

\begin{figure}[htbp]
	\begin{centering}
	\includegraphics[width=0.46\textwidth]{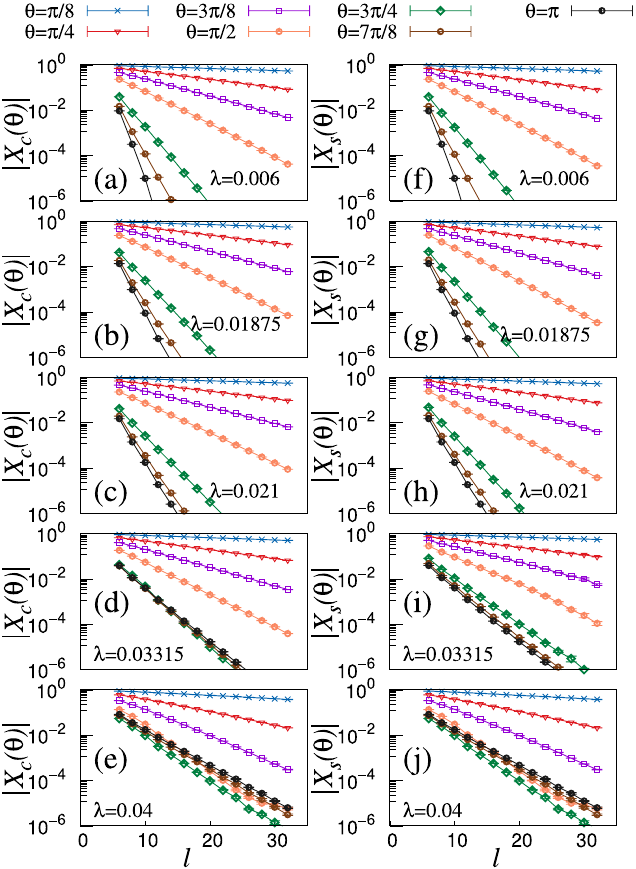}
	\par\end{centering}
	\caption{Disorder operator $X_{c}(\theta)$(a)-(e) and $X_s(\theta)$(f)-(j) obtained for the  model of Eq.~(1) in the main text as function of perimeter $l$ on the  honeycomb lattice.  We consider a  logarithmic  y-axis  and  $L=\beta=15$.   Different subplots correspond to different coupling constants $\lambda$. In each subplot, we present the disorder operator obtained from different rotation angle $\theta$ for comparison.}
	\label{fig:x-vs-lam}
\end{figure}

\begin{figure}[tb]
\begin{centering}
\includegraphics[width=0.48\textwidth]{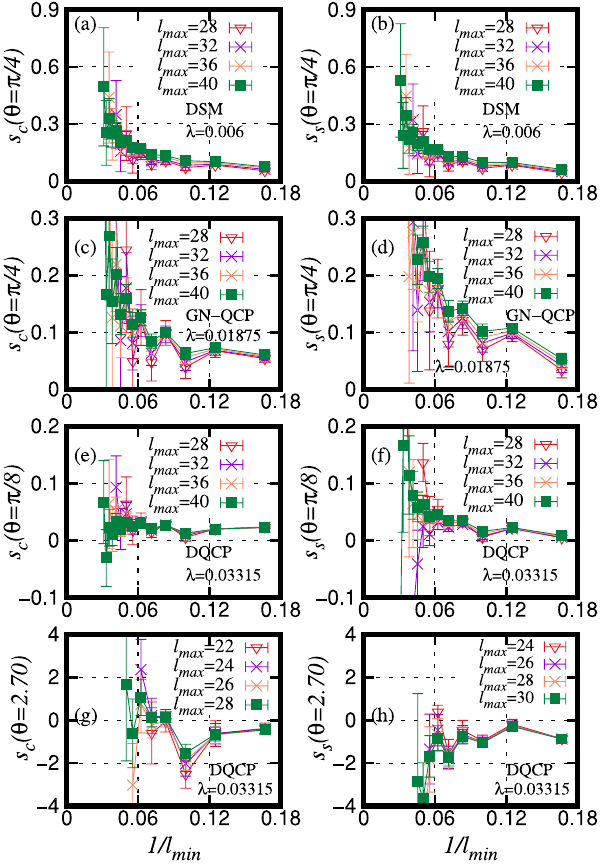}
\par\end{centering}
\caption{Logarithmic coefficient $s(\theta)$ as function of smallest perimeter $l_{min}$ consider in fitting. We use the QMC results of $L=\beta=18$ as the input in numerical fitting.}
\label{fig:fit_window_qmc}
\end{figure}

\begin{figure}[htbp]
	\begin{centering}
	\includegraphics[width=0.46\textwidth]{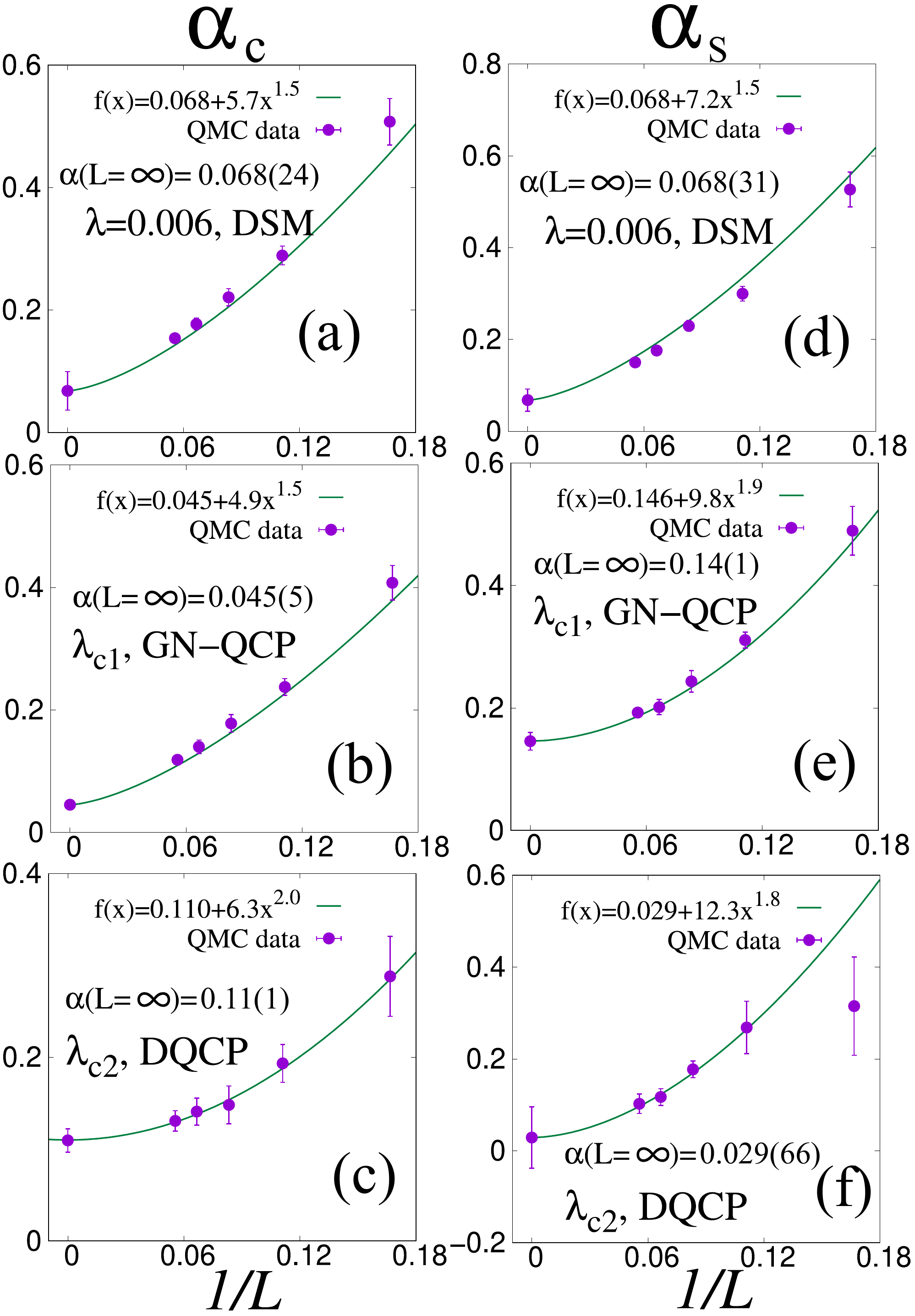}
	\par\end{centering}
	\caption{Coefficient $\alpha_{c}(L)$ (a)-(c) and $\alpha_{s}(L)$ (d)-(f) as function of $1/L$ in the QMC simulation. Different subplot correspond to different coupling constants $\lambda$. In each subplot we use the power law function $f(x)$ to extract the coefficient $\alpha_{c/s}(L=\infty)$ in thermodynamic limit.}
	\label{fig:fig4}
\end{figure}

We now focus on the angle dependence for a given coupling. In Fig.~\ref{fig:x-vs-lam}, we set the coupling constant to be deep in the three phases and also the vicinity of the two critical points and monitor the trends of $X_{c/s}(\theta)$ as function of $\theta$. On the whole, the decay rate of $X_{c/s}(\theta)$ are also dominated by the area law and increase as the angle grows. 

\begin{table}[tbp]
    \caption{Results of coefficient $\alpha(L)$ in thermal dynamic limit by using polynomial fitting in Fig.~\ref{fig:fig4}. } 
    \centering 
    \def\arraystretch{1.5}
    \begin{tabular*}{\linewidth}{@{\extracolsep{\fill} } l l c c c c} 
    \hline\hline 
    \multicolumn{4}{c}{\hspace*{-3em}$\alpha(L=\infty)$ as function of $\lambda$   \hspace*{-3em}} \\
    \hline
    Channel & $\lambda=0.006$ & $\lambda_{c1}=0.01875$ & $\lambda_{c2}=0.03315$ \\[0.5ex]
    \hline\hline
    $\alpha_c(\infty)$ & 0.068(24) & 0.045(5) & 0.11(1) \\ 
	\hline
	$\alpha_s(\infty)$ & 0.068(31) & 0.14(1) & 0.029(66) \\
	\hline\hline
    \end{tabular*}
    \label{table:fit_alpha} 
\end{table}

In Fig.~\ref{fig:fit_window_qmc}   we  present   the  analysis of the  data  carried out  so as to  extract  the logarithmic  correction  to  the  area  law.      As  for  the mean-field  case  we  define  a  fit  window,  and  vary the  bounds   so as  to obtain   reliable  results.  Clearly,  this  is  a delicate  quantity  to  extract,  but  we   have   found  that we  obtain  consistent  results  for   the  fitting  range  $l_{min} \simeq 10$ for  our  largest  $L=18$  system. For the results presented in the main text, we choose $l_{min}=10$ for all system size at DQCP $\lambda=0.03315$ and $l_{min}=8$ for the other coupling constant. The upper bound of the fitting range $l_{max}=2L+4$ for rotation angle $\theta<2.00$ and $l_{max}=\text{min}{(2L+4,28)}$ for large rotation angle $\theta>2.00$.   

As supplementary data, we provide the details of the finite size extrapolation analysis of the quadratic coefficients $\alpha_{c/s}(L)$ in different coupling $\lambda$. To extract the quantity $\alpha_{c/s}(L)$ for a specific system size, we perform numerical fitting using the form $s_{c/s}(\theta)\sim \alpha_{c/s}\theta^2$. We constrain the fitting range of the rotation angle to $\theta\in[0,0.4]$ at DQCP and $\theta\in[0,0.9]$ at other coupling constants. We assume the finite size coefficient obey the power law dependence $\alpha_{c/s}(L)\sim \alpha(\infty)+\kappa L^{-e}$ and using this form in numerical fitting to obtain the true coefficient $\alpha(\infty)$ in the thermodynamic limit. We consider the system size with $L=6,9,12,15,18$ and $\beta=L$ scaling in anticipation of $z=1$. These results are shown in Fig.~\ref{fig:fig4} and Table~\ref{table:fit_alpha}.

\section{$\pi$-flux model}
\label{sec:secVI}
In this section, we begin with the introduction of the $\pi$-flux model. Then we discuss the free case of $\pi$-flux model, including the exact expression for  the density correlation function, the disorder operator, and compare with the 2d DQCP model, which is defined on the honeycomb lattice. To generalize to the interacting case, we investigate $\alpha$   for  the  free case, at  the  GN-Ising QCP    so as  to    obtain an estimate of the current central charge $C_J$  for  this  QCP.  

\subsection{Introduction of $\pi$-flux model }

To probe a   Gross-Neveu   Ising  transition,  form   DSM to QSH , we adopt another fermionic model, $H = H_{\text{f}} + H_{\text{Ising}} + H_{\text{int}}$, defined on the $\pi$-flux lattice.    The  interaction,  $H_{\text{int}}$  describes  the coupling  of  the free Dirac fermions to  a   transverse Ising model,   $H_{\text{Ising}}$.    Specifically, 
\begin{equation}
	\begin{aligned}
		H_{\text{f}} &= -t\sum_{\left \langle ij \right \rangle \sigma} (e^{i \sigma \phi}\hat c_{i \sigma }^\dagger \hat c_{j \sigma } + e^{-i\sigma\phi} \hat c_{j \sigma}^\dagger \hat c_{i \sigma} ) \\
		H_{\text{Ising}} &= -J\sum_{\left \langle pq \right \rangle } \hat s_p^z \hat s_q^z-h\sum_{p} \hat s_p^x \\
		H_{\text{int}} &= \xi \sum_{\langle\langle ij \rangle\rangle \sigma } \hat s_p^z (\hat c_{i\sigma}^\dagger \hat c_{j\sigma} + \hat c_{j\sigma}^\dagger \hat c_{i\sigma})
	\end{aligned}
	\label{eq:Ham_Int}
\end{equation}
$H_{\text{f}}$ describes the nearest neighbor hopping for fermions, $t=1$,  on the $\pi$-flux lattice,  $\phi=\frac{\pi}{4}$, and we request spin-up and spin-down fermions to carrier opposite flux patterns to preserve the time-reversal symmetry  for  the  full  Hamiltonian. $H_{\text{Ising}}$ describes a ferromagnetic $J=1$ transverse-field Ising model. $H_{\text{int}}$ couples the Ising spins with the next nearest neighbor fermion hoppings.  The coupling constant $\xi = \pm 1$ has a staggered sign structure alternating between neighboring plaquettes, i.e., $+ (-)$ for solid (dashed) bonds as illustrated in Fig. 1(e) in the main text.  Tuning the transverse field $h$ the model undergoes  a GN-Ising QCP at $h_c=4.11$   from a   DSM phase (at $h< h_c)$   and a QSH state (at $h>h_c$). Fig. 1(b), (e) are the corresponding phase diagram and the choice of the entanglement region $M$. 

\subsection{Density correlation function in the free case}

Here we investigate  another lattice regularization  of  Dirac fermions,  namely  the $\pi$-flux model.  The Hamiltonian reads, 
\begin{equation}
	\begin{aligned}
		H_{\text{f}} = &-t_1\sum_{\left \langle ij \right \rangle} (e^{i \phi}\hat c_{i }^\dagger \hat c_{j } + e^{-i\phi} \hat c_{j}^\dagger \hat c_{i} ) \\
		&-t_2\sum_{\left \langle\langle ij \right \rangle\rangle} (e^{i \phi}\hat c_{i }^\dagger \hat c_{j } + e^{-i\phi} \hat c_{j}^\dagger \hat c_{i} )
	\end{aligned}
	\label{eq:Ham}
\end{equation}
The sketch of the  $\pi$-flux model is shown in Fig.~1(e) in the main text. The fermions are  located on the lattice  sites, colored in green and blue, indicating two sublattices. $\hat c^{\dagger}, \hat c$ are fermion creation and annihilation operators. The fermion hopping term between nearest two green and blue sites has an extra phase factor $e^{i\phi}$, whose sign is positive (negative) along the direction of the  arrow.  The  choice $\phi=\frac{\pi}{4}$  produces a  phase $\pi$ in  each  plaquette. $t_1$ represents nearest hopping for different sublattices, while $t_2$ represents the next-nearest hopping along diagonal lines.
The Hamiltonian gives rise to two Dirac cones located at $(0,\pi)$ and $(\pi,0)$~\cite{he2018Dynamical,liuDesigner2020}. As mentioned, we set $t_1=1, t_2=0$. 
The low energy physics of  the  $\pi$-flux model   is   equivalent to  that of the honeycomb lattice  and, in the absence  of spin degree  of freedom, 
  is  described by two  two component Dirac spinors~\cite{Ihrig2018Critical}.  Furthermore, $N_f$ is directly related to the current central charge in CFT. For single free Dirac fermion, $C_{J,free}=2$. In the language of lattice model, e.g. both $\pi$-flux model, and honeycomb lattice, there are two Dirac points in the Brillouin zone, such  that $N_f=2C_{J,free}$. 
Note that the Hamiltonian in Eq.~\eqref{eq:Ham} involves  spinless fermions. When one considers the limit   $\xi=0$ the two spin flavor decouple, corresponding  to  $N_f=8$.

To verify $\alpha=\frac{N_f}{(4\pi)^2}$ at small angle for Dirac cones~\cite{lliesiuBootstrapping2018},  we calculate the density correlation function in $\pi$-flux model,
\begin{equation}
	D_{\pi}(\mathbf{r})=\langle \hat n_{\mathbf{r}_i} \hat n_{\mathbf{r}_j} \rangle - \langle \hat n_{\mathbf{r}_i} \rangle \langle \hat n_{\mathbf{r}_j} \rangle = \frac{N_{F}}{(4 \pi)^{2}r^{4}} \frac{1-2 \frac{x y}{r^{2}}(-1)^{x+y}}{2}
	\label{eq:den_corr}
\end{equation}
where, $\mathbf{r} = \mathbf{r}_i - \mathbf{r}_j$, $r$ is the absolute distance of $\mathbf{r}$, and $x(y)$ is the component of $\mathbf{r}$ along the  x(y) direction.   Here   corrections  to  the   rotationally  invariant  IR  result  are   taken  into  account.   
To exhibit the above expression in the lattice model, we provide analytic results of the density correlation function in Fig.~\ref{fig:den_corr}. In  lattice models, such as the $\pi$-flux model~\cite{he2018Dynamical,liuDesigner2020}, the  density correlation function has small variance with respect to Eq.~\eqref{eq:den_corr}. The red line indicates the 
$\frac{N_F}{(4\pi)^2 r^4}$  form where $N_F=4$ for the $\pi$-flux model.  In Fig.~\ref{fig:den_corr} (a), we plot $D(\mathbf{r})$ for each site at $L=1000, t_1=1, t_2=0$  and compare with the analytical formula, yellow dots. 
Differences are apparent
 at small $r$, corresponding large momentum, or short wave length contributions. The short wave length  properties are determined by the lattice structure or the microscopic interacting coefficients. Next in Fig.~\ref{fig:den_corr} (b), we change $t_2=0.1$ to verify this difference, where two Dirac cones still remain. We find the difference between yellow and blue dots to  be at  small  $r$. To further explore the lattice model implementation, we plot the various system size in Fig.~\ref{fig:den_corr} (c) and discover when $r$ is comparable with the system size, $D(\mathbf{r})$ will obviously deviate from the $\frac{1}{r^4}$ behavior, which is regarded as a finite system effect. 
 Fig.~\ref{fig:den_corr} (d) extracts the lattice sites on the diagonal line, i.e., $x=y$, where the oscillation term in Eq.~\eqref{eq:den_corr} gives 1. 
 Here  we  note  that  the lattice  constant  is  set  to unity  such  that $x$ and $y$     can be half-integers  or  integers. 
 The  results   nicely  match  with the red line  at moderate $r$ for each system size. Thus we conclude that  the density correlation function calculated in the lattice model differs   from  the continuum limit both at small $r$ and at  $r \sim L$,  due to lattice microscopic details and finite system sizes, respectively. 

\begin{figure}[htp!]
	\begin{minipage}[htbp]{0.49\columnwidth}
		\centering
		\includegraphics[width=\columnwidth]{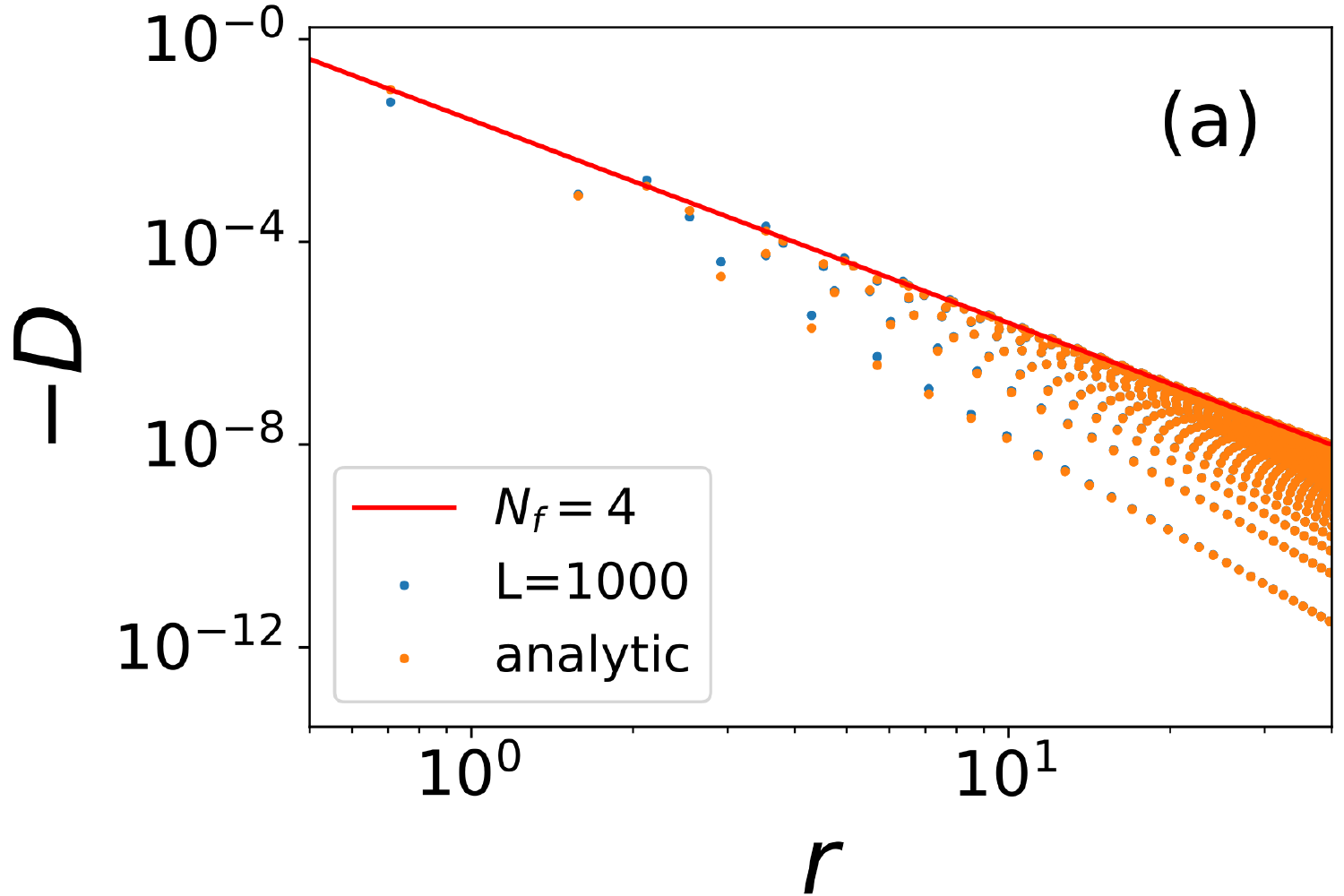}
	\end{minipage}
	\begin{minipage}[htbp]{0.49\columnwidth}
		\centering
		\includegraphics[width=\columnwidth]{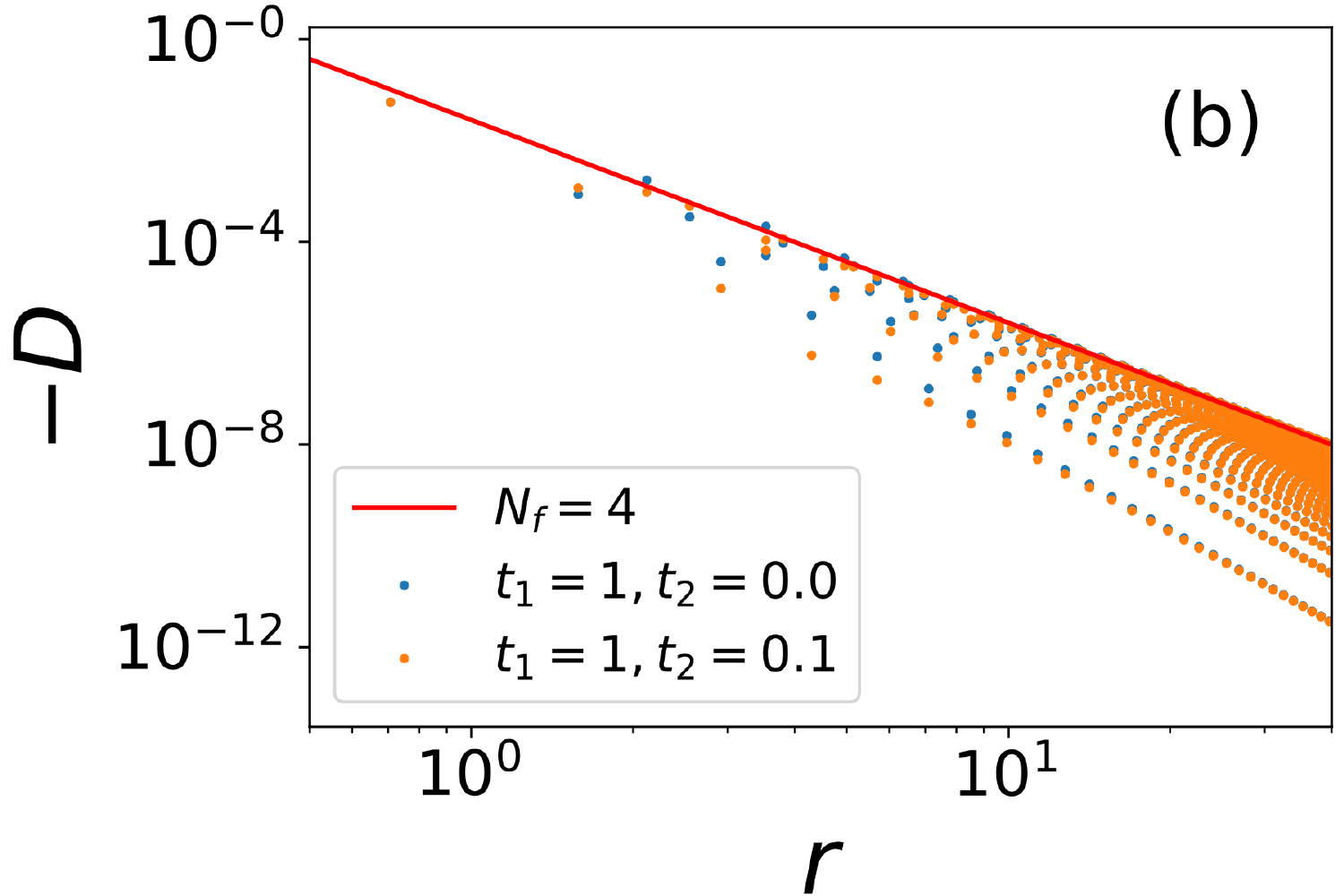}
	\end{minipage}
	\begin{minipage}[htbp]{0.49\columnwidth}
		\centering
		\includegraphics[width=\columnwidth]{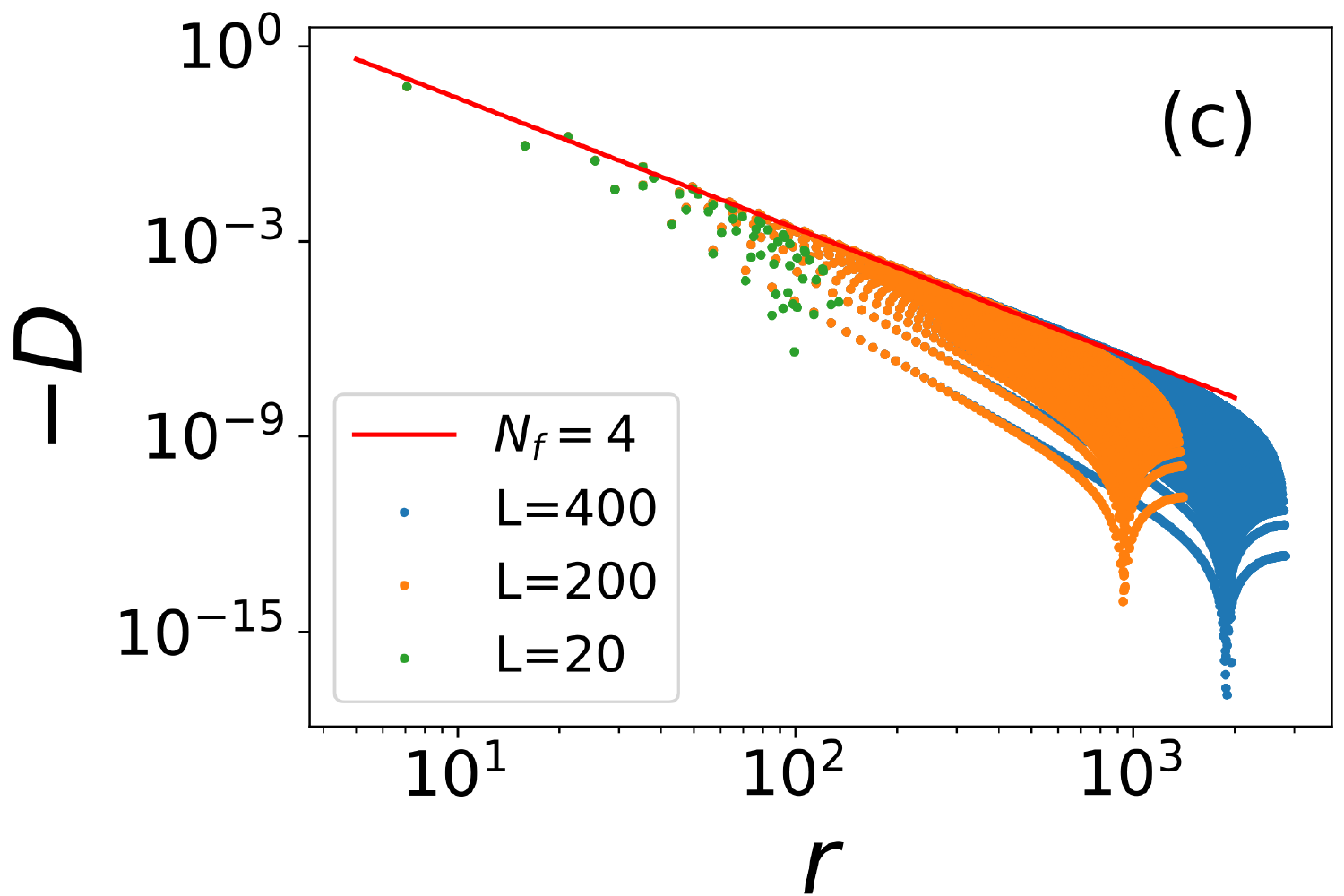}
	\end{minipage}
	\begin{minipage}[htbp]{0.49\columnwidth}
		\centering
		\includegraphics[width=\columnwidth]{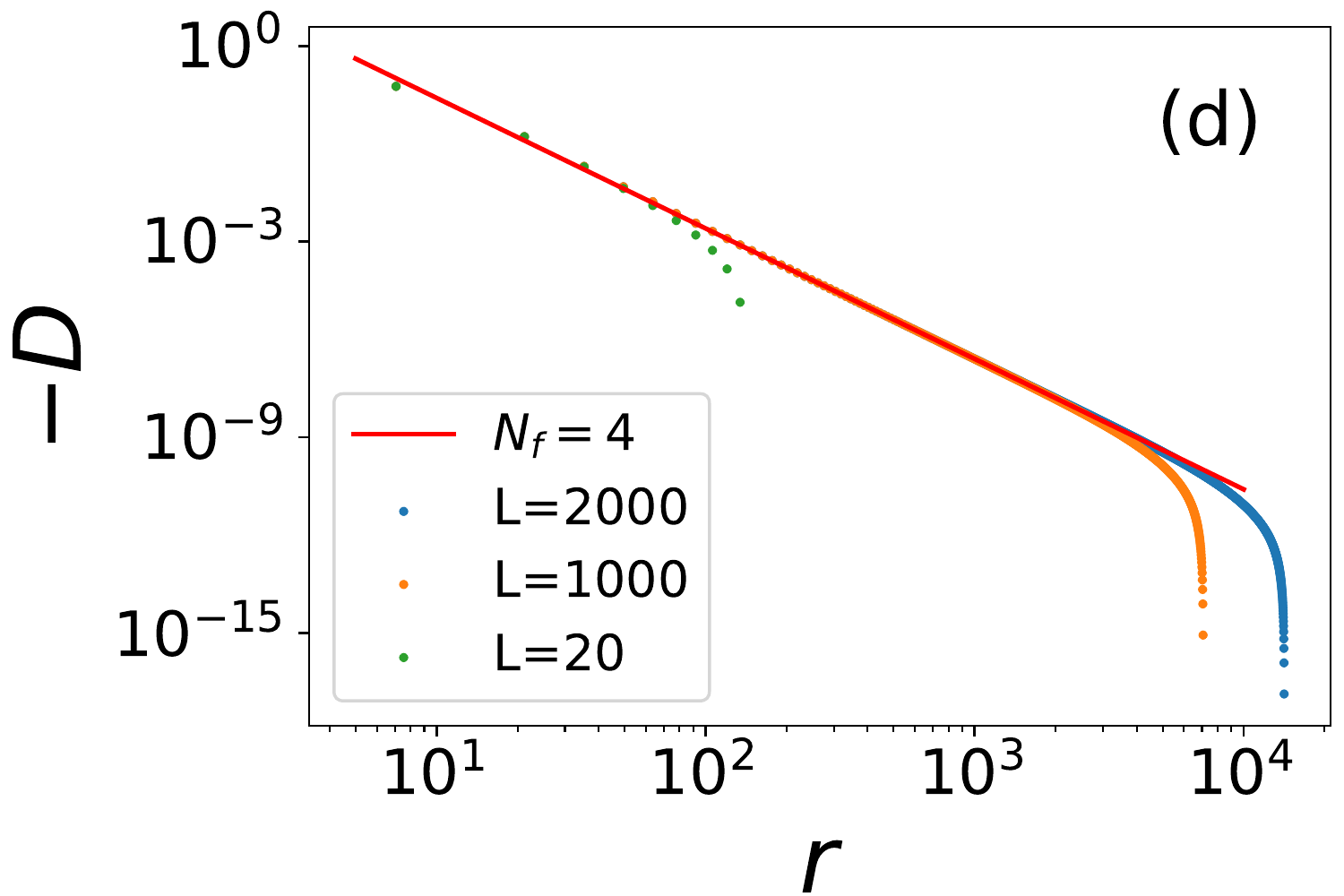}
	\end{minipage}
	\caption{Analytical density correlation function in Eq.~\eqref{eq:den_corr} versus $r$ compared with lattice model results on a ln-ln scale. The red line in each subplots is the same functional form $\frac{N_F}{(4\pi)^2 r^4}$, where $N_F=4$ for the spinless $\pi$-flux model.  The  points with same color and $r$ represent various   values of  $x,y$. (a) $D(r)$ for $L=1000, t_1=1, t_2=0$ and dots for Eq.~\eqref{eq:den_corr}, plotted in blue and yellow. (b) $D(r)$ for $L=1000, t_1=1, t_2=0.0$ and $L=1000, t_1=1, t_2=0.1$, plotted in blue and yellow. (c) $D(r)$ for  $t_1=1, t_2=0.0$ at various system $L=400,200$ and $20$ plotted in blue, yellow and green. (d) $D(r)$ for $t_1=1, t_2=0.0$ at diagonal line, i.e. $x=y$ with various system $L=2000, 1000$ and $20$ plotted in blue, yellow and green.}
	\label{fig:den_corr}
\end{figure}

\subsection{The disorder operator in the free case} 
Ultiziling Eq.~\eqref{eq:den_corr}, we calculate the disorder operator at small $\theta$ in the free case,
\begin{equation}
	\frac{\ln(|X(\theta)|)}{\theta^2} = -a_1 l + \frac{N_f}{(4 \pi)^2} \ln l + a_0
	\label{eq:dis_ope}
\end{equation} 
The coefficient of area law $a_1$ is determined by the lattice parameters, e.g., $t_1, t_2$. The logarithmic correction  is universal,
 and only depends on $N_f$. Following, we will focus on the subleading logarithmic correction.

\begin{figure}[htp!]
	\begin{minipage}[htbp]{0.49\columnwidth}
		\centering
		\includegraphics[width=\columnwidth]{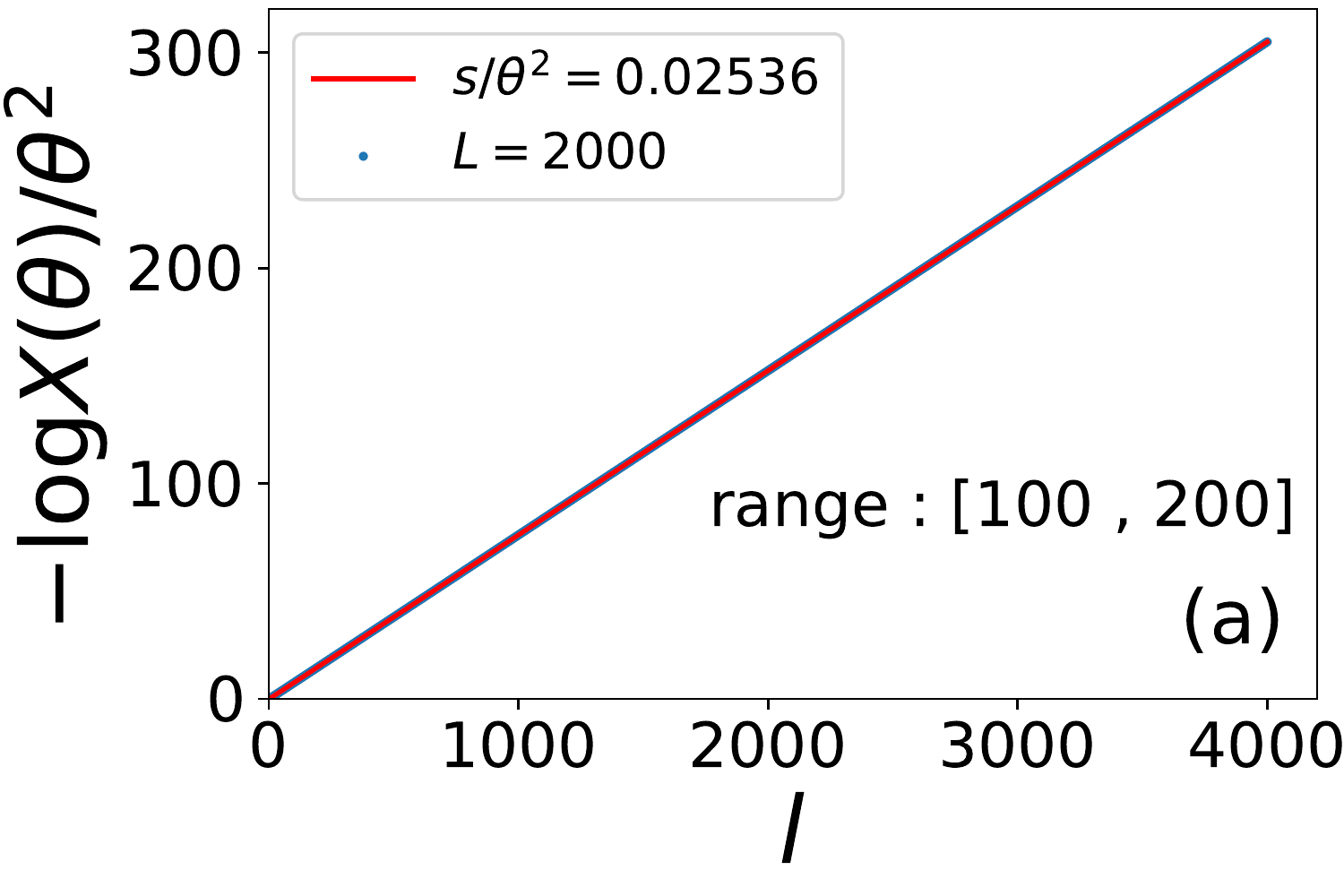}
	\end{minipage}
	\begin{minipage}[htbp]{0.49\columnwidth}
		\centering
		\includegraphics[width=\columnwidth]{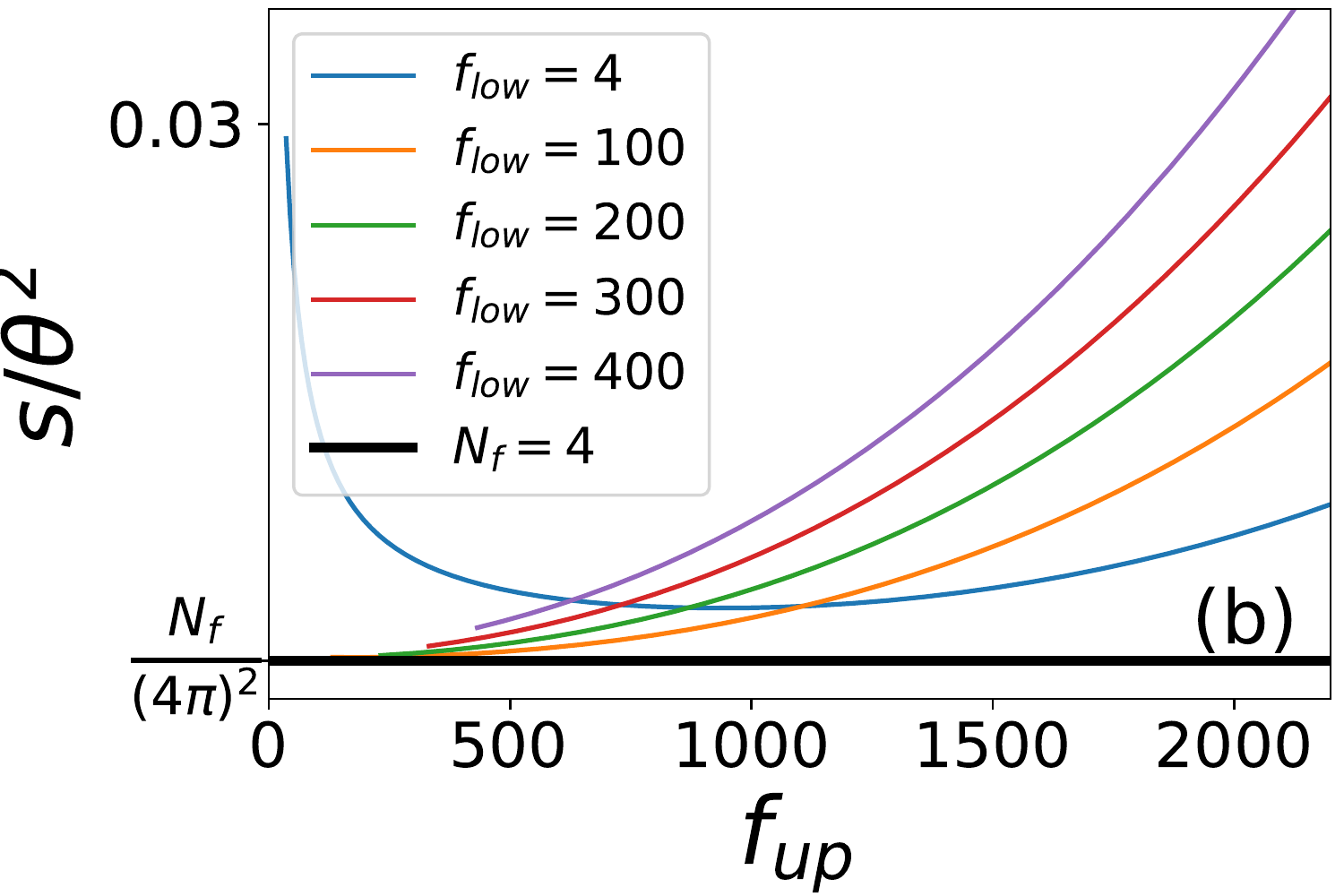}
	\end{minipage}
	\begin{minipage}[htbp]{0.49\columnwidth}
		\centering
		\includegraphics[width=\columnwidth]{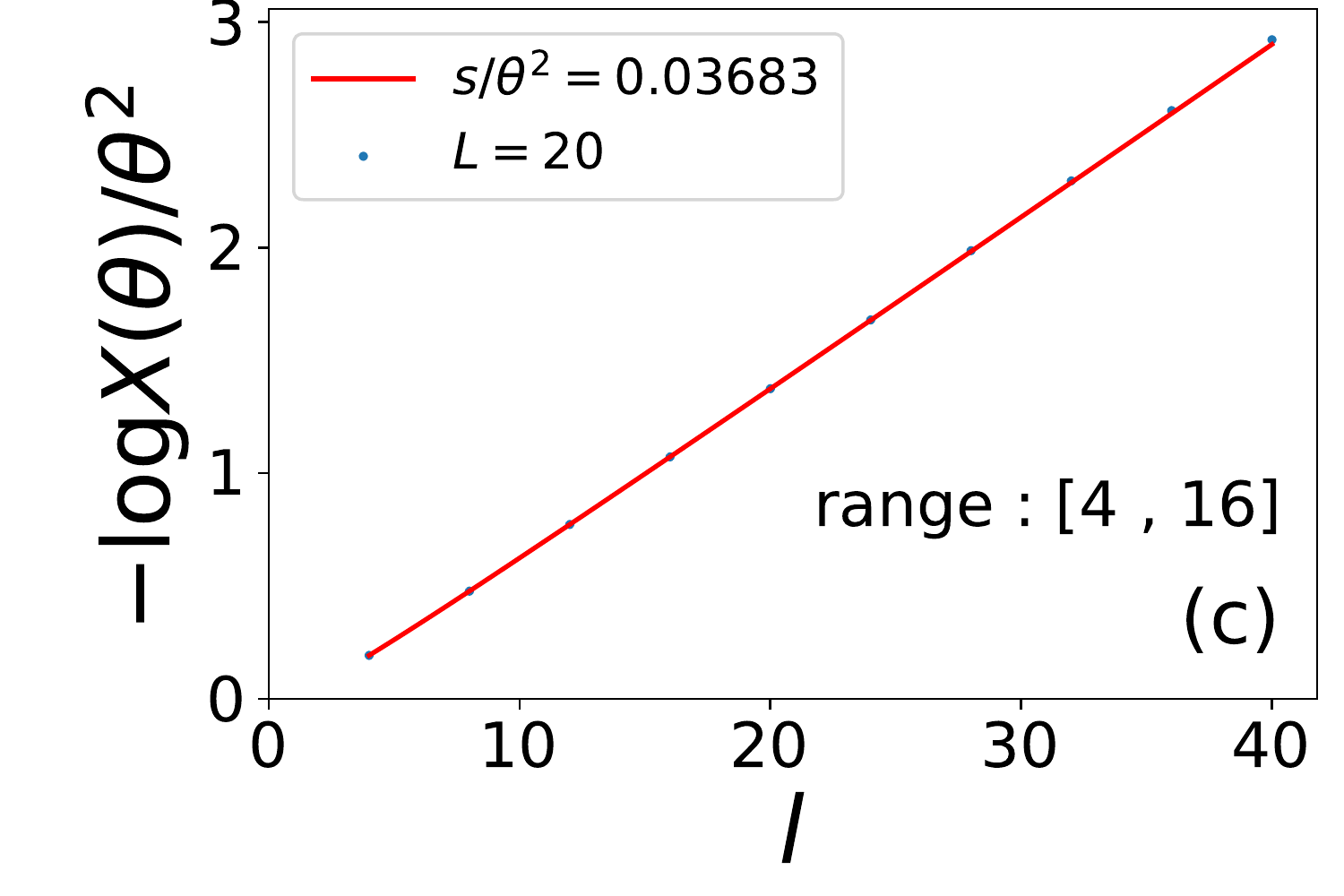}
	\end{minipage}
	\begin{minipage}[htbp]{0.49\columnwidth}
		\centering
		\includegraphics[width=\columnwidth]{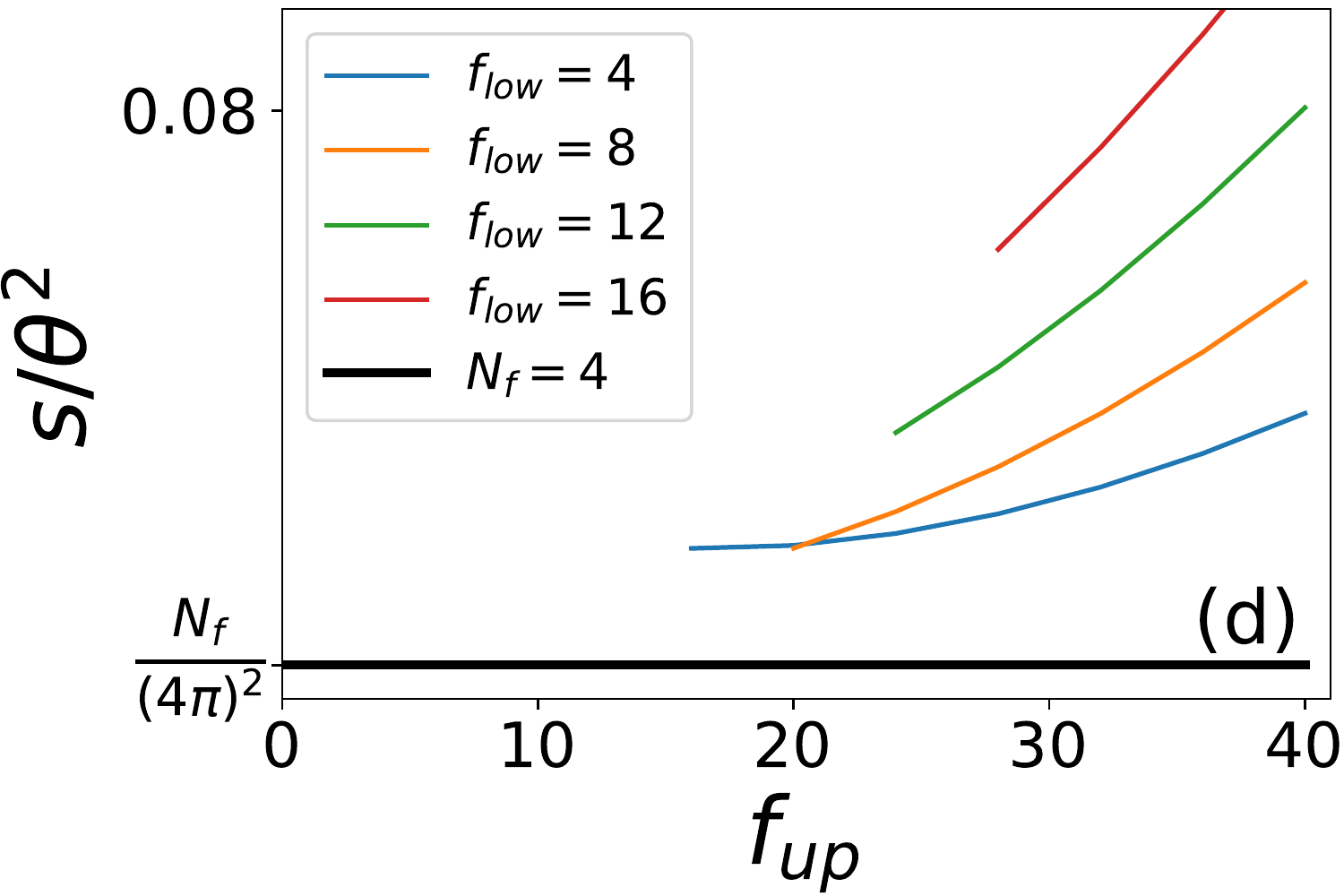}
	\end{minipage}
	\caption{ The scaling behavior of the disorder operator on the $\pi$-flux lattice at small angle $\theta=0.1$ and choice of fitting range for determining  coefficient $s$ to show the finite size effect. (a,b) for $L = 2000$, (c,d) for $L=20$, comparable with the size of the interacting system. (a,c) plots the original disorder operator data versus perimeter $l$, where $-\ln X$ seems like one linear function. (b,d) show $s(\theta)/\theta^2$ versus various fit range $l \in \left[ f_{low}, f_{up}\right]$. The analytic value at thermodynamic limit $\frac{N_f}{4\pi^2}$ is plotted by black solid line. We find if the lower boundary $f_{low}$ is too small, as the same order of $1$, e.g. the blue line in (b), the fit results will overestimate the value. On the contrast, when the upper boundary $f_{up}$ gets larger, the fit results gradually deviate the analytic value. Thus we conclude the proper choice to get $s(\theta)/\theta^2 = \frac{N_f}{4\pi^2}$ is $1 \ll f_{low} < f_{up} \ll L$. For small size $L=20$ in (d), we find the condition is hard to satisfy, thus leads to overestimated value for $s$.}
	\label{fig:diso_ope}
\end{figure}

Taking small $\theta$ as example, Eq.~\eqref{eq:dis_ope} is strictly valid only in the limit  $L, l \rightarrow \infty$, since one needs to compute the integral over Eq.~\eqref{eq:den_corr} in region $M$, i.e. $\int_{\mathbf{r}_1 \in M} d^2 \mathbf{r}_1 \int_{\mathbf{r}_2 \in M} d^2 \mathbf{r}_2 D_{\pi}(\mathbf{r}_1 - \mathbf{r}_2)$. Techinically, in Monte Carlo simulation for interacting fermionic model, $l$ and $L$ are both finite. Since $\ln l$ is small compared to  the leading $l$ term,  at finite system size, it is  hard to  obtain $\alpha$ in the thermodynamic limit by direct fitting for the function form Eq.~(1) in the main text at current system size, i.e. $L=18$ for interacting system.  Compared with previous studies  on the Bose-Hubbard model, the disorder operator also possess  sub-leading $\ln l$  corrections~\cite{wangScaling2021}. We numerically find that  the coefficient $-a_1$ of the leading area law term in free Dirac fermions  is much larger than that in the  Bose-Hubbard model, 
shown in Fig.~\ref{fig:diso_ope}(a) for $L=2000$ and (c) for $L=20$
which leads to difficulties to extract  $\alpha$. 
Besides, we notice that the choice of the  fitting range  changes the  result,  Fig.~\ref{fig:diso_ope}. We denote the fit range as $l \in \left[ f_{low}, f_{up}\right]$. We plot several choices of $f_{low}$ as curves with various colors, the  x-axis being $f_{up}$.  From  the above analysis  we conclude 
 severe finite size effect for the coefficient.  

To obtain $\alpha$ in the thermodynamic limit, we adopt two  strategies. In the main text pertaining  to interacting systems, we directly fit the log correction for each system size, and then extrapolate to $L \rightarrow \infty$. Both errorbars  stemming from  the Monte Carlo data as  well as  the  
systematic  error generated  by  variation by the choice of the fitting  range are  taken into account  for  the  estimate of $\alpha(L)$. In contrary to the interacting case, for free systems, we are able to reach  large enough  lattices  so as   to  approach the thermodynamic limit.  By changing the fitting range, we show that one can  obtain  approximate values of  $\alpha$  in the large  system size limit.  In the following, we will show how to obtain the value, corresponding to $N_f=4$.

To carry out  the analysis, we introduce  the concept of  so-called optimal fit range. In the finite size, we observe that all choices  of  the  fitting 
 range will overestimate the log correction  as compared to  the  value in the  thermodynamic limit. Therefore for  a  given  system size, we 
 consider the fitting  range, which gives the smallest log correction. We find that this smallest  log correction  will also gradually approach the thermodynamic limit value as $L$ increases. 

In Fig.~\ref{fig:diso_ope} (b) at a large system size, we observe if $1 \ll f_{low} < f_{up} \ll L$ is satisfied, for example $f_{low} = 100, f_{up} = 200$, the fit result $\frac{s}{\theta^2}=0.02536$ in (a) is close to the analytic value  in the  continuum  limit $\frac{N_f}{(4\pi)^2}=0.02533$. For comparison, at $L=20$ in (c,d), we show that whatever the choice of fit range, the result will be larger than  $\frac{N_f}{(4\pi)^2}$. Even the optimal fit range $\left[ 4, 16\right]$, corresponding to the closest value to the thermodynamic limit at  the  considered  system size, gives $\frac{s}{\theta^2}=0.03683$, almost more than one-half  the value in  the thermodynamic limit. To conclude, we can extract  the coefficient as a function of system size $s(\theta,L)$ with the optimal fitting  range  strategy  and  then extrapolate  to  the  thermodynamic limit: $\frac{s(\theta \rightarrow 0, L  \rightarrow \infty)}{\theta^2} = \frac{N_f}{(4\pi)^2}$.  As we showed in Fig.~\ref{fig:diso_ope}, the most optimal value is always the minimum among all possible fit range. This criterion is more unambiguous than above mentioned $1 \ll f_{low} < f_{up} \ll L$. Hence, one may expect $s(\theta,L)$ to  gradually approach $s(\theta \rightarrow 0,\infty)$ as $L$ increase.

\subsection{Comparison between the $\pi$-flux model and honeycomb lattice implementation}
\label{sec:secVII}

\begin{figure}[htp!]
	\begin{minipage}[htbp]{0.49\columnwidth}
		\centering
		\includegraphics[width=\columnwidth]{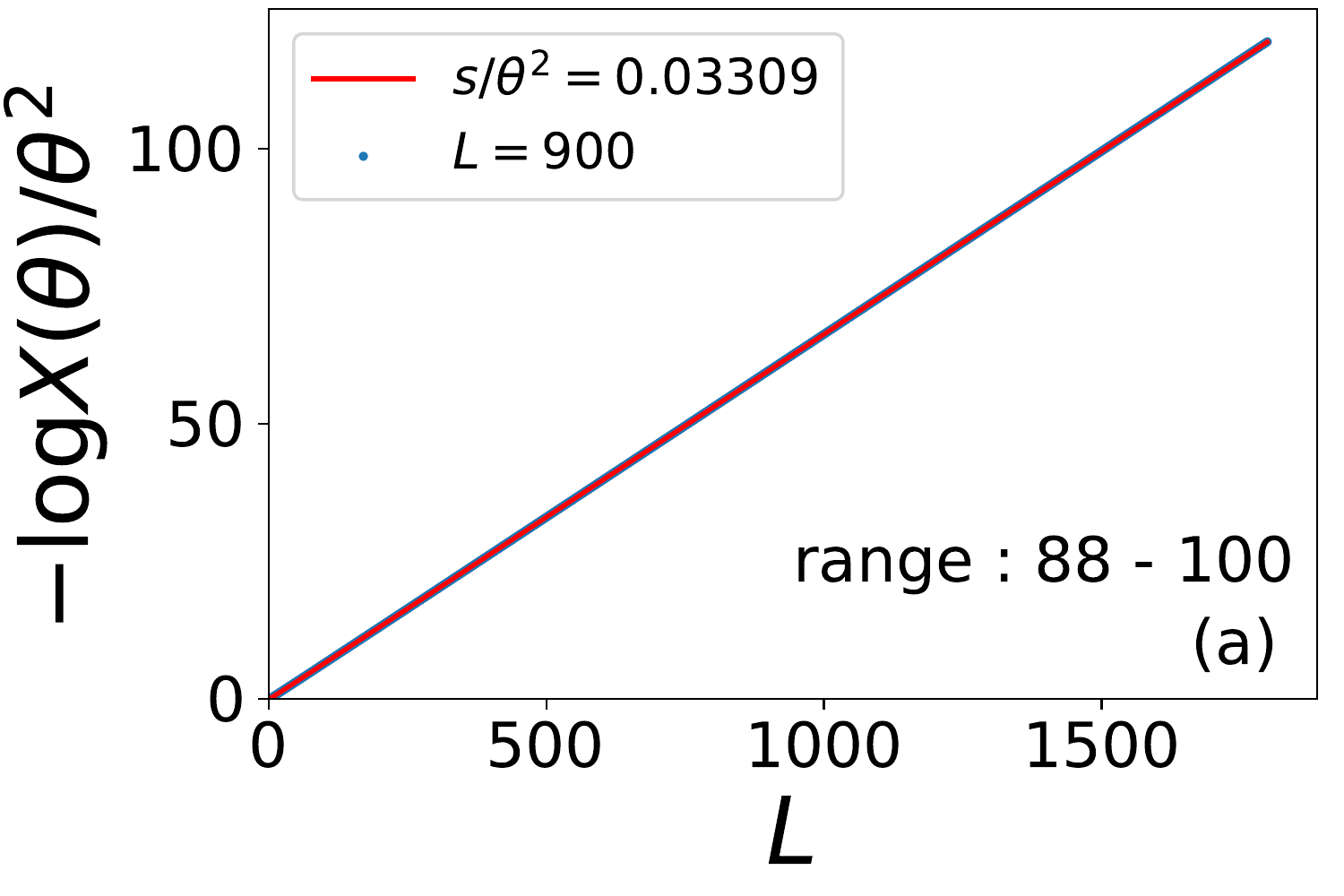}
	\end{minipage}
	\begin{minipage}[htbp]{0.49\columnwidth}
		\centering
		\includegraphics[width=\columnwidth]{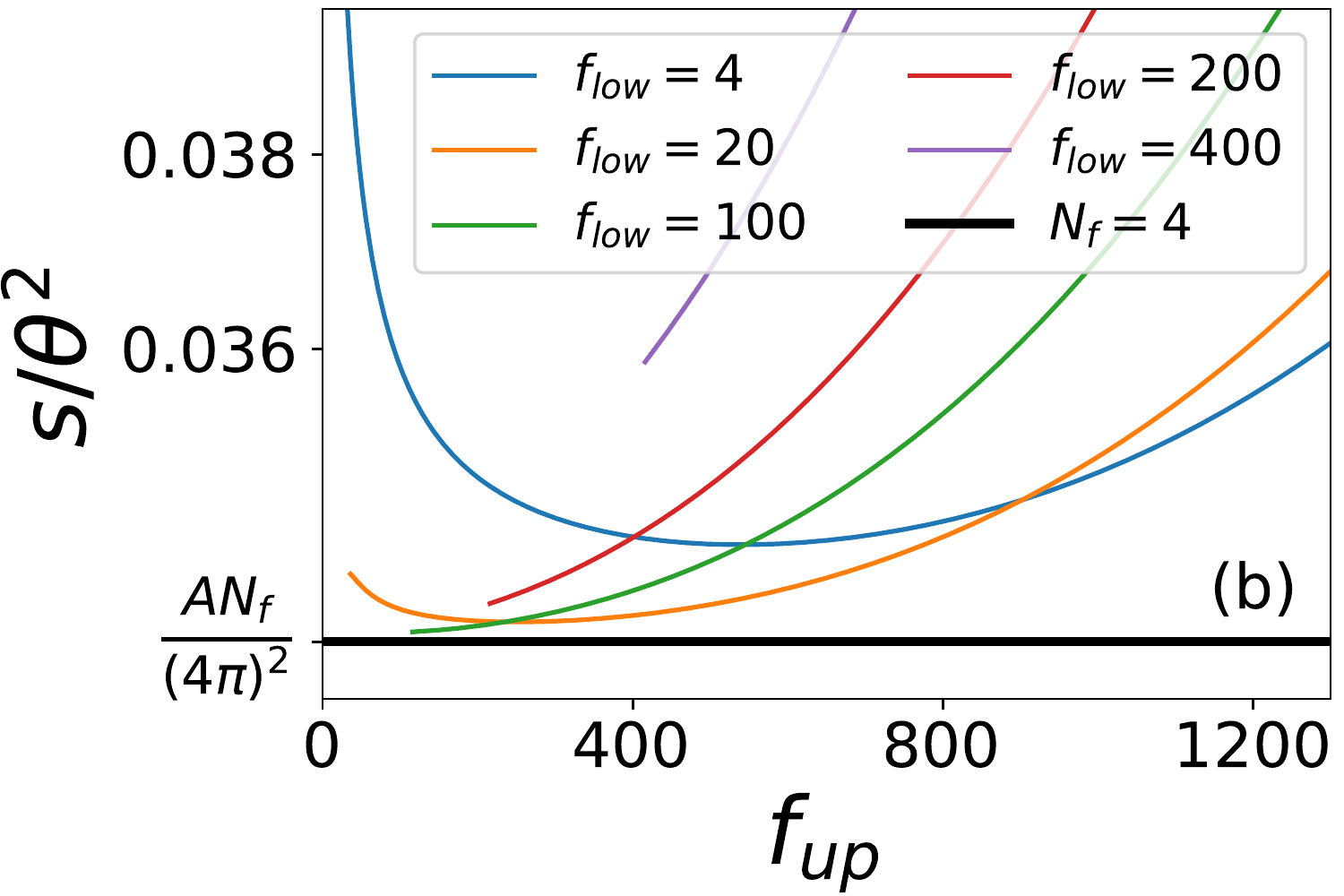}
	\end{minipage}
	\begin{minipage}[htbp]{0.49\columnwidth}
		\centering
		\includegraphics[width=\columnwidth]{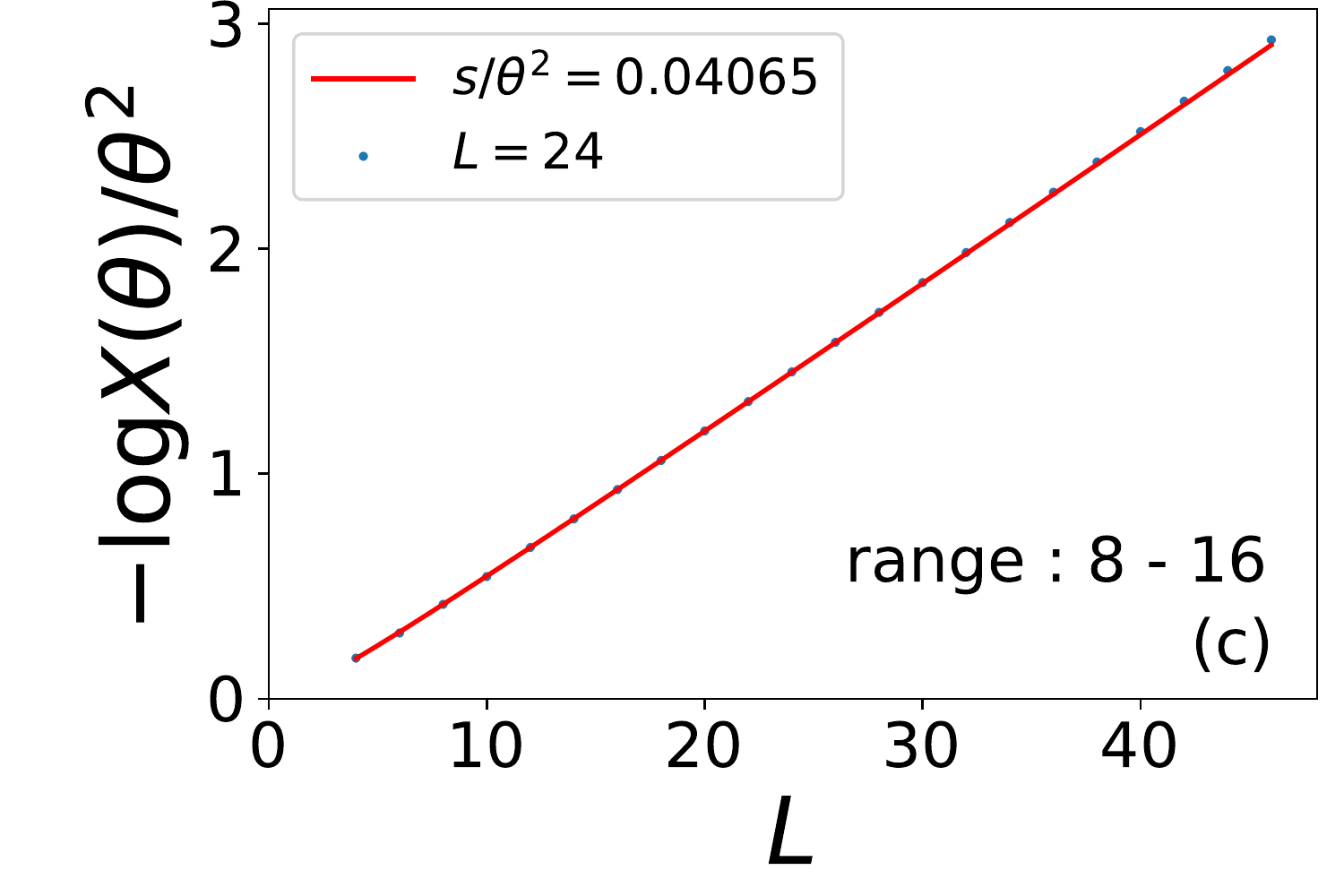}
	\end{minipage}
	\begin{minipage}[htbp]{0.49\columnwidth}
		\centering
		\includegraphics[width=\columnwidth]{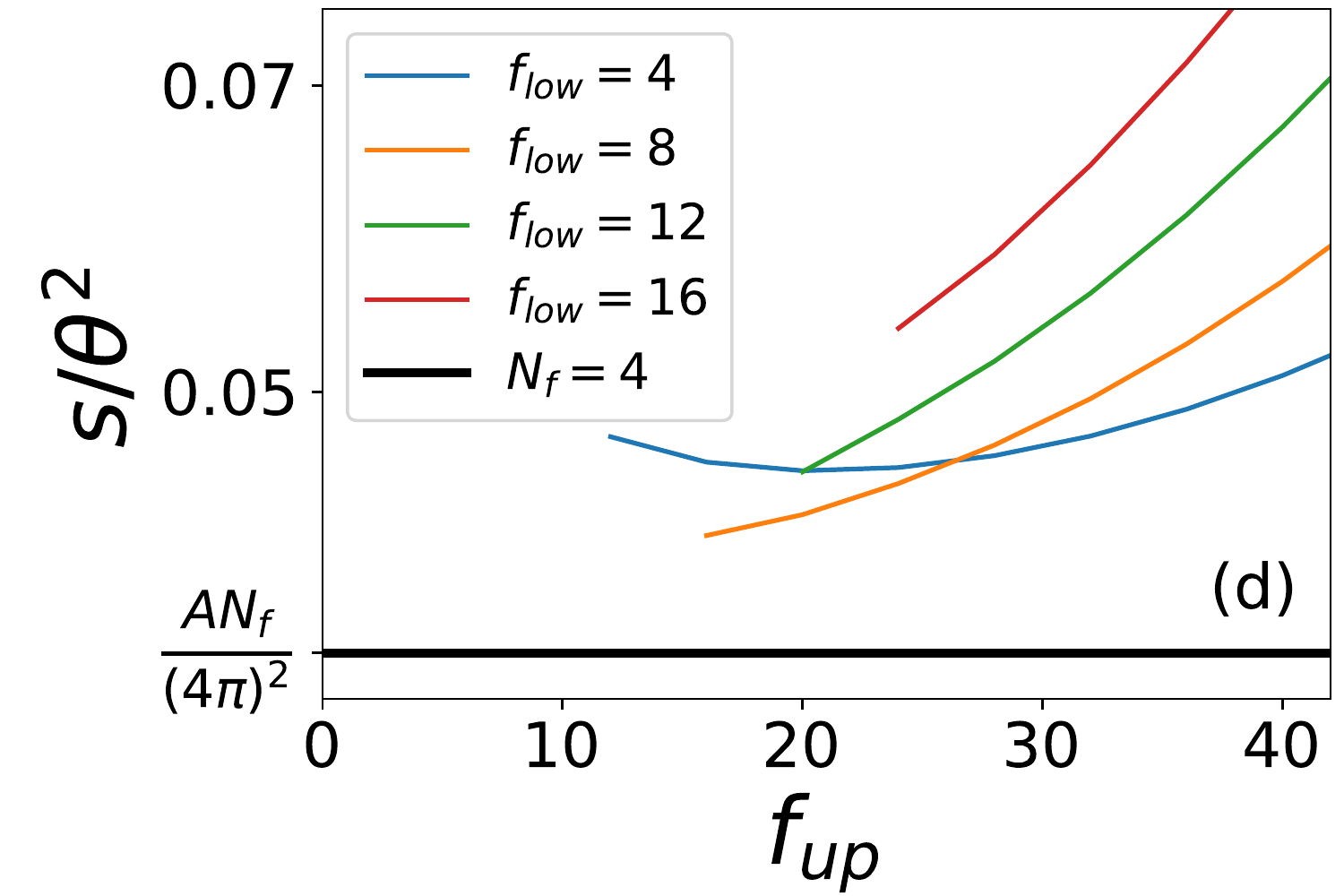}
	\end{minipage}
	\caption{The scaling behavior of the disorder operator on the honeycomb lattice at small angle $\theta=0.1$ and choice of fitting 
	 range to  determine the  coefficient $s$. (a,b) for $L = 900$, (c,d) for $L=24$, comparable with the size of the interacting system. (a,c) plots the original disorder operator data versus perimeter $l$, where from  $-\ln X$ it seems  hard to extract the log correction. 
	 (b,d) shows $s(\theta)/\theta^2$ versus $f_{up}$ for  various fit ranges $l \in \left[ f_{low}, f_{up}\right]$. The analytic value in the  thermodynamic limit $\frac{N_f}{4\pi^2}$ is plotted by a black solid line. We find  that 
	if the lower bound  $f_{low}$ is too small,
	blue line in (b), the fit results will overestimate the value.  In  contrast, when the upper boundary $f_{up}$ gets larger, the fit results gradually deviates
	from  the analytic value. Thus we conclude  that the  proper choice to obtain  $s(\theta)/\theta^2 = \frac{N_f}{4\pi^2}$ is for  $1 \ll f_{low} < f_{up} \ll L$. For small sizes $L=24$ in (d), we find the condition is hard to satisfy, thus leading to overestimated values for $s$.  }
	\label{fig:diso_ope2}
\end{figure}

As  for the $\pi$-flux model, free fermions on the honeycomb lattice  form Dirac cones, located at the   $(\frac{2\pi}{3},\frac{2\pi}{3\sqrt{3}})$ and $(-\frac{2\pi}{3},\frac{2\pi}{3\sqrt{3}})$  points  in the  Brillouin zone.   This  corresponds to 
$C_{J,free}=2$ and $N_f=4$. We analytically  solve the density correlation  function to obtain: 
\begin{equation}
	D_h(\mathbf{r})=\langle \hat n_{\mathbf{r}_i} \hat n_{\mathbf{r}_j} \rangle - \langle \hat n_{\mathbf{r}_i} \rangle \langle \hat n_{\mathbf{r}_j} \rangle \sim \frac{N_{F}}{(4 \pi)^{2}r^{4}}
	\label{eq:free4}
\end{equation}
$D_h(\mathbf{r})$ has  a similar  from to   $D_{\pi}(\mathbf{r})$ aside  from  the oscillation term, shown in Fig.~\ref{fig:den_corr}, indicating  that both 
have  the  same scaling behavior of the disorder operator $al + b\ln l +c$   when carrying out  the integral over  $M$  of  the  density fluctuations.  
It must be emphasized that here the region $M$ is a  parallelogram, whose degree of corners are $60^{\circ}, 60^{\circ}, 120^{\circ}, 120^{\circ}$. Here, we use the conclusions in Ref.~\cite{wu2021universal}, in which the contribution by corners is described by $s \propto \sum_{\alpha} f(\alpha)=\sum_{\alpha} 2(1+(\pi-\alpha)\cot(\alpha))$, where the summation runs over the  interior angle for region $M$. Thus, we obtain the modification due to the   geometry of  the parallelogon $M$-region  over  the  square  of  $A \approx 1.30$.  Hence we  have: 
\begin{equation}
	\frac{\ln(|X(\theta)|)}{\theta^2} = -a_1 l + \frac{AN_f}{(4 \pi)^2} \ln l + a_0
	\label{eq:free5}
\end{equation} 
In Fig.~\ref{fig:diso_ope2}  we carry  out  a similar  analysis as for the  square  lattice,   Fig.~\ref{fig:diso_ope}. We find $\frac{s(\theta \rightarrow 0, L=24)}{\theta^2} = 0.04065$, $\frac{s(\theta \rightarrow 0, L=900)}{\theta^2} = 0.03309$, and $\frac{s(\theta \rightarrow 0, L=24)}{\theta^2} = 0.03299$. The difference for $L=900$ and $L=24$ compared with the thermodynamic limit value is $0.3 \%$ and $23\%$.

The result can be applied to 2d DQCP model at $\lambda < \lambda_{c1}$, where one expects in thermodynamic limit with $A \approx 1.3$, $N_f=8$, $\alpha = \frac{AN_f}{(4\pi)^2} \approx 0.0658$.



\subsection{$C_J$ at Gross-Neveu QCP}

In this subsection, we   discuss  the results at Gross-Neveu QCP, and compare the disorder operator at the  
 Gross-Neveu-Ising and Gross-Neveu-Heisenberg transitions. 
 As  mentioned in the main text, the disorder operator at small $\theta$ also obeys the expression for Eq.~(1) in the main text, where the $\ln l$,  coefficient  named $s(\theta) \sim \alpha \theta^2$ has the relation with the current central charge $C_J$ in the  corresponding  
 CFT. Theoretically, one expects $\alpha = \frac{AN_{\sigma}C_J}{8\pi^2}$   at the   Gross-Neveu QCP, where the angle modification coefficient $A = 1$ for the  $(90^{\circ},90^{\circ},90^{\circ},90^{\circ})$ square region and $A \approx 1.30$ for $(60^{\circ},60^{\circ},120^{\circ},120^{\circ})$ for  the  parallelogon. We aim to calculate $C_J$ for the Gross-Neveu Ising  and  Heisenberg  QCPs. 
  However, the system size in the interacting case is limited at $L = 20$ for $\pi$-flux, and $L=24$ for honeycomb, leads to the deviation from the analytic value by fitting, as shown in Fig.~\ref{fig:diso_ope}(d) and Fig.~\ref{fig:diso_ope2}(d). That directly indicates that the choice of fit range brings non-negligible influences on the value of $\alpha$.  
  Besides, the original data of interacting case has errorbars, which differs with the non-interacting case, as shown in Fig.~\ref{fig:diso_ope} and Fig.~\ref{fig:diso_ope2}. Thus, it is  difficulty to fit the disorder operator to obtain $C_J$. Here, based on the knowledge that the disorder operator is the integral over  the  density correlation function $D(\mathbf{r})$, we directly calculate $D(\mathbf{r})$ by DQMC simulations and compare the result 
  by  fitting the disorder operator.  We display the calculations in the interacting Hamiltonian in Eq.~\eqref{eq:Ham_Int}   in the main text, corresponding   to $A=1$, and Gross-Neveu-Ising universality.   Fig.~\ref{fig:intdirac} shows the original data for $D(\mathbf{r})$ in  a  $\ln$-$\ln$ scale for   our  largest   system size. At   first sight, $D(\mathbf{r})$ at $h_c$ still obeys the  $\frac{1}{r^4}$ relation and has a small difference with 
   the free results. To investigate the difference carefully, we plot the ratio between $D(\mathbf{r})$  and $D_{free}(\mathbf{r})$ for various system sizes. At small $r$, the ratio seems to be independent of $L$, and gradually decreases. We regard this as the effect on the lattice microscopic details, which is not universal. Conversely, at large $r$, the ratio which exceeds 1 results from the finite size effect, as $r$ becomes comparable to $L$. At moderate $r$, we find a minimum for each $L$, and we expect  that a plateau corresponding to  the minimum will appear at large $L$. We thus plot the so-called extrapolation line in pink and construct  the  plateau as we expect and finally estimate the value of $\frac{D}{D_{free}} \sim 0.77$  in  thermodynamic limit.

\begin{figure}[htp!]
	\centering
	\includegraphics[width=\columnwidth]{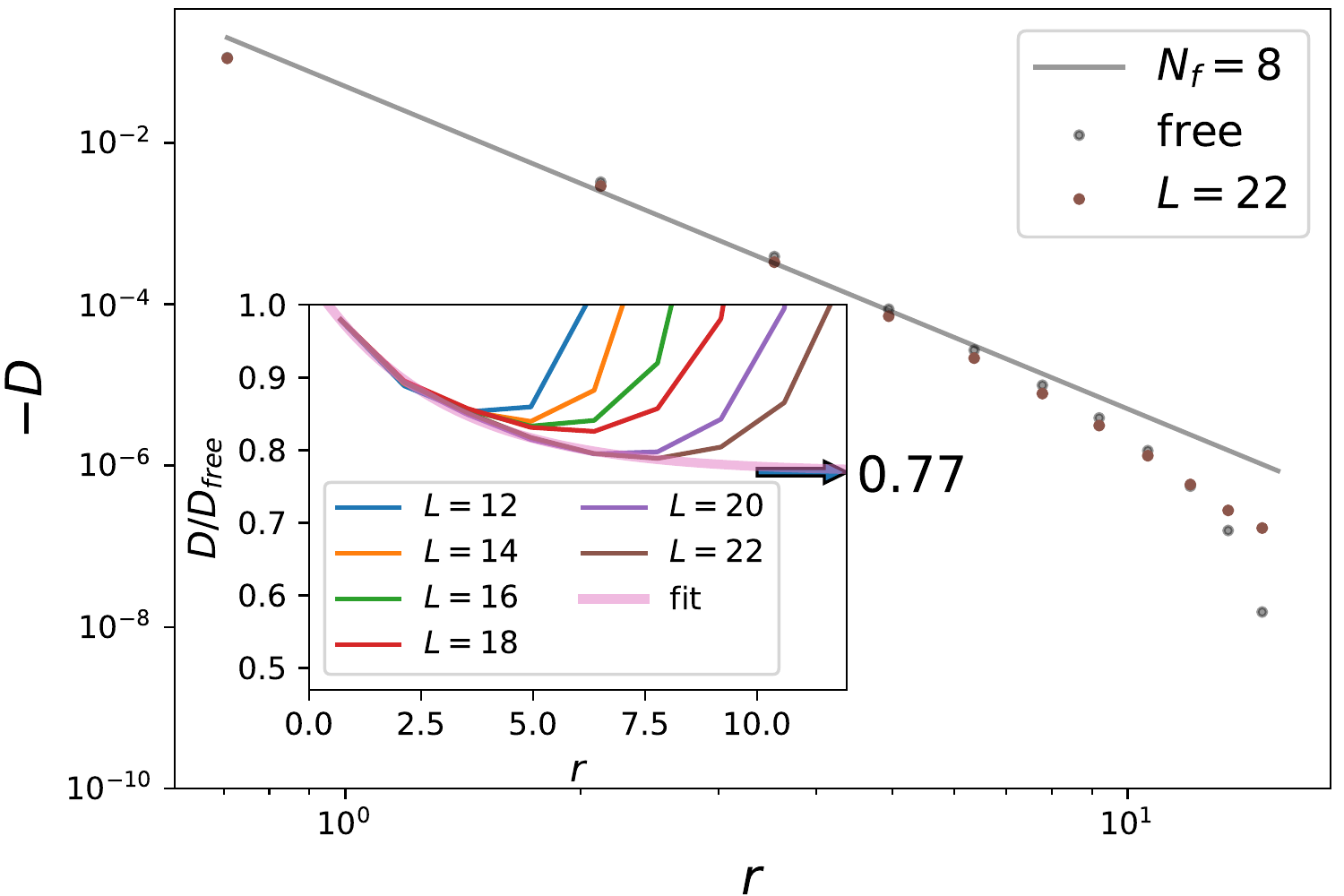}
	\caption{Density correlation function $D(\mathbf{r})$ versus $\mathbf{r}$  for  the  Gross-Neveu Ising QCP. We directly plot $D(\mathbf{r})$ along the  diagonal for our  largest system size, e.g. $\mathbf{r}_x=\mathbf{r}_y$, and compared with  to the free case $D_{free}$, plotted in grey on a  $\ln$-$\ln$ scale. Obviously both follow the $\frac{1}{r^4}$ behavior. The difference in such scale is small and hard to distinguish. To obtain $C_J$,  $\frac{D}{D_{free}}$ is displayed in  the  inset  for various  values of  $L$. We expect at $1 \ll r \ll L$, $\frac{D}{D_{free}}$ to converge to the thermodynamic limit. Limited by the system size, we use the  grey thick line to depict the extrapolated value with respect the existing data. We fit for  several data points along the envelop   with the form,  $y=c+a e^{-bx}$. We estimate $\frac{D}{D_{free}} \equiv c \sim 0.77$ in the  thermodynamic limit.}
	\label{fig:intdirac}
\end{figure}

\section{1d DQCP model}
\label{sec:1ddqcp}

In this section, we provide data on the convergence check of DMRG simulations.

\begin{figure}[htp!]
	\centering
	\includegraphics[width=\columnwidth]{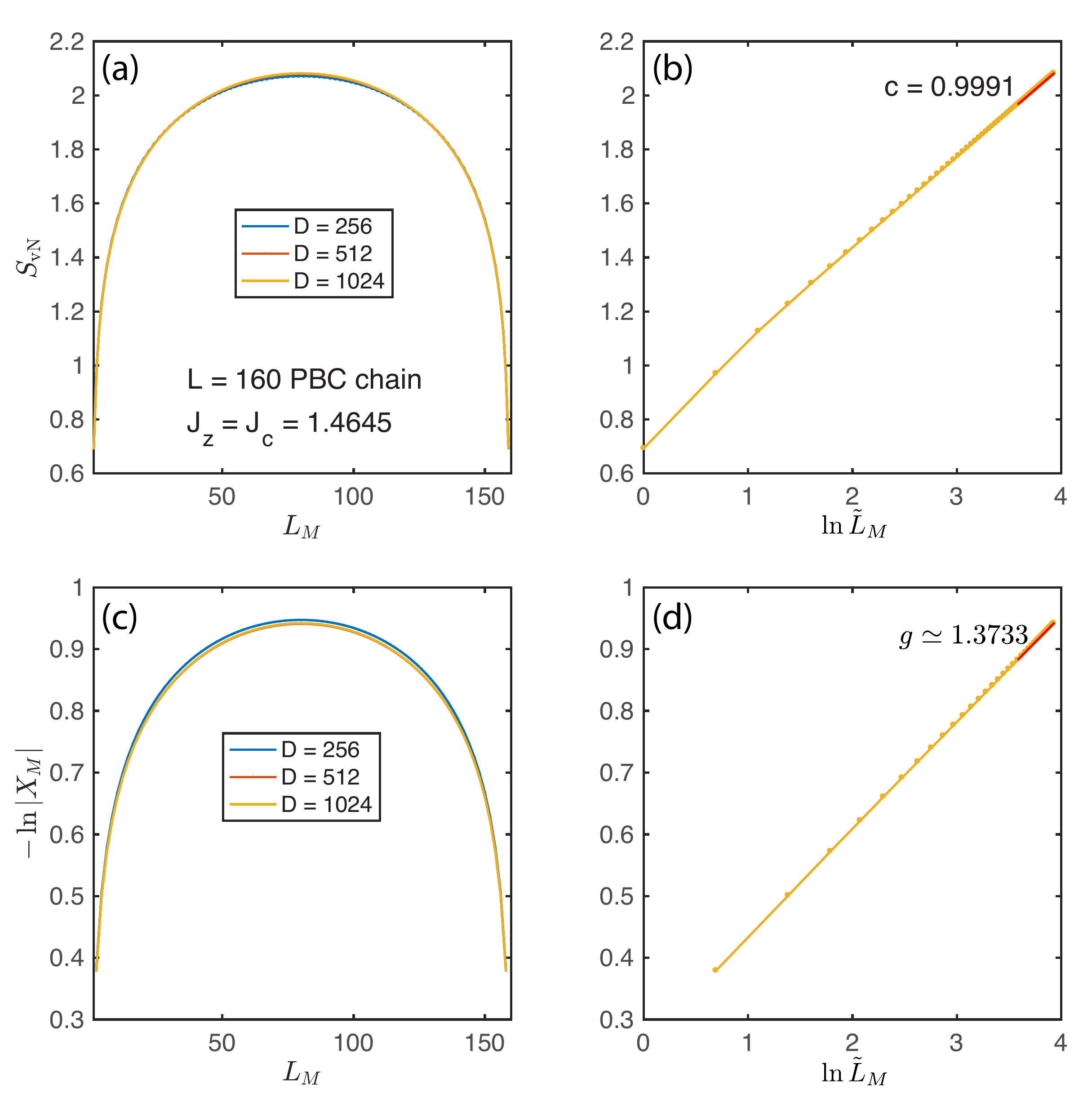}
	\caption{ We  consider  the $L=160$ 1D DQCP model at the  DQCP point $J_z=J_c=1.4645$ 
	with periodic boundary condition (PBC). 
	(a, b)   shows  the entanglement (von Neumann) entropy $S_\mathrm{vN}$ and 
	(c, d) the logarithmic disorder operator  
	$-\ln|X_M|$ for the   DMRG ground states at  different bond dimensions 
	$D=256, 512, 1024$.  
	$L_M$   corresponds  to the  subsystem size and $\tilde L_M = \frac{L}{\pi}\sin\frac{\pi L_M}{L}$.
	is  the conformal distance 
	Here, the color code for different bond dimension $D$ as indicated in both 
	the panel (a) and (c). 
	From (b) and (d), the central charge $c=0.9991$ and Luttinger parameter 
	$g=1.3733$ are extracted from the rather straight line composed by the 
	largest 40 data (as indicated by the red solid line).  
	  }
	\label{fig:dmrgconv}
\end{figure}

Throughout our DMRG simulations for different system size $L=64, 96, 128, 160$, 
we keep up to 1024 bond states, which render a small truncation error of 
$\delta \rho < 5\times10^{-9}$. In this section, we show that  for  our  largest system size 
$L=160$ and at the  critical point $J_z=J_c=1.4645$, 
the data presented in the main text is  well   converged.

In Fig.~\ref{fig:dmrgconv} (a) and (b), we show the entanglement (von Neumann) entropy 
$S_\mathrm{vN}=-\mathrm{tr}(\rho_M \ln \rho_M)$ as functions of subsystem size $L_M$
 and the corresponding conformal distance $\tilde L_M = \frac{L}{\pi}\sin\frac{\pi L_M}{L}$. 
 Here $\rho_M = \mathrm{tr}_M |\psi\rangle\langle\psi|$ is the reduced density matrix 
 obtained   by tracing out the degrees of freedom in subsystem $M$. 
 From panel (a) we can see that, $S_\mathrm{vN}$  is well converged for $D\geq 512$.
 We then plot the $D=1024$ data versus the logarithmic conformal distance $\ln{\tilde L_M}$
 following the expected CFT behaviour $S_\mathrm{vN} =\tfrac{c}{3} \ln{\tilde L_M}$, 
 from which $c=0.9991$ is extracted. 
 
For the disorder operator $X_M = \prod_{i\in M} \sigma^z_i$, we repeat the similar analysis 
in Fig.~\ref{fig:dmrgconv}(c) and (d). Again, the logarithmic disorder operator 
$-\ln{|X_M|}$ is  well converged for $D\geq512$. 
From $-\ln{|X_M|} = \tfrac{g}{8} \ln{\tilde L_M}$, $g=1.3733$ is extracted at the 1D DQCP point.

\clearpage



\end{document}